\documentclass[11pt]{article}

\usepackage{graphicx}  

\usepackage{fullpage}      
\usepackage{color}         

\newcommand{\deriv}[2]{\ensuremath{\frac{d#1}{d#2}}  }       

\newcommand{\pderiv}[2]{\ensuremath{\frac{\partial#1}{\partial#2}}  } 

\newcommand{\minifderiv}[2]{\ensuremath{\delta#1/\delta#2}  }   

\renewcommand{\exp}[1]{e^{#1}}

\newcommand{\vect}[1]{\ensuremath{\mathbf{#1}}}

\newcommand{\epseff}{\ensuremath{\epsilon_{e}} } 
\newcommand{\epseffbar}{\tilde{\epsilon_{e}} }   

\newcommand{\Action}{\ensuremath{\mathcal{F}_{\Omega}}} 
\newcommand{\action}{\ensuremath{f_{\Omega} } }         

\newcommand{\fig}[1]{Fig. \ref{#1}}  
\newcommand{\tab}[1]{Table \ref{#1}} 

\newcommand{\sect}[1]{Section \ref{#1}}   
\newcommand{\appdx}[1]{Appendix \ref{#1}} 

\newcommand{\eqn}[1]{Equation (\ref{#1})}  
\newcommand{\eq}[1]{(\ref{#1})}            

\newcommand{\ut}[1]{\,\rm{#1}}  

\begin{document}

\title{Nonadditivity of Polymeric and Charged Surface Interactions: Consequences for
  Doped Lamellar Phases.}

\author{O. A. Croze and M. E. Cates\\\\ 
School of Physics, University of Edinburgh\\
JCMB King's Buildings, Mayfield Road\\
Edinburgh EH9 3JZ, Scotland}

\maketitle

\begin{abstract}
We explore theoretically the modifications to the interactions between charged surfaces
across an ionic solution caused by the presence of dielectric polymers. Although the chains are neutral, the 
polymer physics and the electrostatics are coupled; the
intra-surface electric fields polarise any low permittivity species (e.g., polymer)
dissolved in a high permittivity solvent (e.g., water). This 
coupling enhances the polymer depletion from the surfaces and increases the 
screening of electrostatic interactions, with respect to a model which treats polymeric
and electrostatic effects as independent. As a result, the range of the ionic contribution to the osmotic interaction between surfaces is decreased, while that of the polymeric contribution is increased. These changes modify the total interaction in a nonadditive manner. Building on the results for parallel surfaces, we investigate the 
effect of this coupling on the phase behaviour of polymer-doped smectics. 
\end{abstract}


\section{Introduction}

Many processes in soft and biological systems take place in water and involve the interaction
of fatty components, such as membranes or macromolecules. The polar nature of the
aqueous environment means these components often acquire surface charges, so that electrostatics plays
a key role in determining their physical behaviour. The subject has undergone a
substantial revival recently, especially because of its biological relevance
\cite{ElecEffSoftMatt}. 

The description of such electrostatic systems generally invokes a continuum approximation
(see, e.g., Kjellander in \cite{ElecEffSoftMatt}):
the electrostatic properties of the solvent (water) and of all uncharged components
are accounted for via their electrical permittivity. Membranes in water, for
example, have been usefully 
modelled as dielectric films of low permittivity residing in a high permittivity medium
 \cite{MenesPincus, NetzDielSlab}. In
general, any dielectric components present, even if neutral, will be involved in 
modulating electrostatic interactions, because of their
polarisation in the electric fields generated by the charged components of the system. 
This creates nonadditivity, which is
not always recognised. Specifically, in modelling neutral polymers between charged surfaces,
the electrostatic and polymeric contributions to the intersurface forces are
usually treated as independent (see, e.g., sec. 10.7 of \cite{Russel}). With charged polymers
(polyelectrolytes or polyampholytes), their effects on fields is often 
attributed solely to the polymer charges, neglecting the dielectric backbone 
\cite{BorukhovAndelman}. An
exception is the work of Khokhlov et al. \cite{ElecEffSoftMatt}, in which the 
polarisation of polyelectrolytes is explicitly included in a description of their
adsorption onto an oppositely charged surface. Dielectric phenomena are also well known to
be implicated in the distribution of electrolytes in the neighborhood of proteins
\cite{Perutz}.   
  
In this paper we directly address the interdependence of the polymer
physics and electrostatics when charged surfaces interact across solutions of neutral polymers. \if{This
coupling cannot be ruled out {\it a priori} because of the dielectric nature of the
polymers (polarised by the surface double layers).}\fi To gauge the extent of the nonadditive
``coupling effects'', we have constructed a simple mean field model which includes such 
coupling, and calculate the force between charged surfaces
across a neutral polymer solution. The model is then adapted to predict the phase behaviour of polymer-doped
lamellar phases. Recent investigations of these composite liquid crystals have shown that
a substantial amount of polymer can be incorporated in the lamellar phase and can induce
phase separation (see \cite{KotzKosmella} and references therein). The polymer has been
found to reside either within the bilayers \cite{Williams1, Williams2} or within the water layers;
in the latter case it may adsorb onto the bilayers \cite{Ficheux95, Ficheux97, Javierre01} or it may not
\cite{LigourePorte93, LigoureTheoryExp97, Javierre01}; this depends on the local polymer--bilayer
interaction. Without charges, the problem 
provides an experimental realisation of the textbook example of a polymer solution
confined in a slit \cite{LigourePorte93, DeGennesBook}. Most lamellar
phases are charged, however. Some existing descriptions
correctly recognise that polymers can affect electrostatic interactions by reducing the
effective permittivity of 
the solution confined between bilayers \cite{Sear1,  Ligoure2000, Linse1,
  KekicheffPEG}, but ignore the feedback of the electrostatics on the 
distribution of polymer segments themselves. 
A fully consistent description 
must either address this, or give a good reason to neglect it. Below we explore this issue further.
For simplicity our work is restricted to the case of nonadsorbing polymers whose
monomer density vanishes at either of the confining walls.

The paper is organised as follows. \sect{ModIISetUpApproxSect} describes our mean field
model and how it can be used to investigate the properties of polymer-doped lamellar
phases. The results of our model are then presented and discussed. In
\sect{ModelIIResults} we show how, for experimentally reasonable parameters, an account of
coupling modifies the variation of 
both the electrostatic potential and the polymer concentration between surfaces of fixed
separation. As a result of these changes, the osmotic interaction between the plates is
also modified with respect to the uncoupled results from the same
model. With the mapping described in \sect{ModIISetUpApproxSect}, the osmotic pressure can be
used as an equation of state to predict the phase behaviour of doped lamellar phases. The
details of this procedure and the results of parameter
variation for the phase behaviour of doped smectics
are presented and discussed in \sect{ResPhasDiags}. \sect{RelRealSystSect} discusses 
the relation to experiment and in
\sect{ModIIConclCommSummSect} we draw our conclusions.

\section{Model \label{ModIISetUpApproxSect}}

\begin{figure}[tbph]
\centering  
\includegraphics[width=0.4\linewidth]{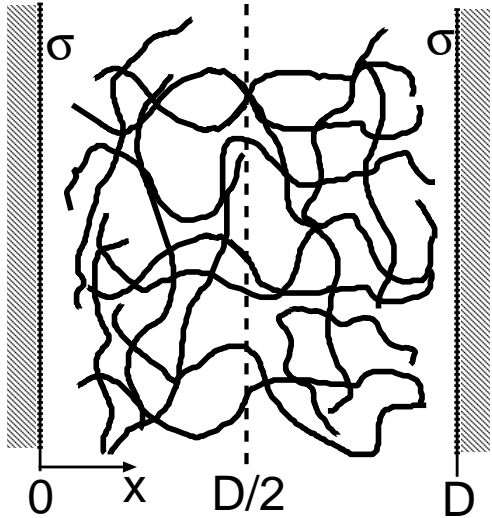} 
\caption{\label{PolymerChain} The situation modelled: an ionic solution of nonadsorbing
  neutral polymers confined between charged surfaces, separated by a distance
  $D$ and bearing a surface charge density $\sigma$. The ions in solution
  with the polymers are not shown for clarity. Also not shown is the electrolyte and
  polymer reservoir with which the system is in equilibrium.}
\end{figure}

We consider  a solution of neutral polymers,
confined between charged surfaces (\fig{PolymerChain}) and contacting a reservoir with
which it can exchange heat, polymer chains and electrolyte (salt). By virtue of the
surface counterions and the salt ions, the solution is electrolytic and screens
the charged surfaces. Since the polymers are nonadsorbing, their monomer density vanishes at both plates.

\subsection{Variational Formulation \label{VarFormSect}}

With these premises, the free energy characterizing the system
is the grand potential $\Omega(\Gamma,T,\mu_i, \mu_p)$, minimised at equilibrium when the solution's volume,
$\Gamma$, its temperature, $T$, and chemical potentials, $\mu_i$ ($i=+,-$) for the ions 
and  $\mu_p$ for  the polymer, are fixed. $\Omega$ comprises energetic and entropic 
contributions from the polymer solution, the ions and electrostatics. Adopting a mean
field approach, such contributions are conveniently expressed in terms of the monomer
volume fraction $\phi(\vect{r})$, the ion number density $n_i(\vect{r})$ and the
electrostatic potential $V(\vect{r})$. As indicated, these variables are expected to vary
with position $\vect{r}$ between the surfaces, so that the free energy needs to be
expressed as a density functional, the null variations of which yield the equations of our
model. Because the electrostatic potential and the ion densities are not independent
variables, but are constrained to obey the Maxwell equation $\nabla \cdot (\epsilon \nabla
V)=-\sum_i n_i q_i$ (where $q_i$ is the ion charge), it is convenient to adopt the
variational formulation of electrostatics \cite{Schwinger} to write the free energy as an
``action'':
\begin{equation}\label{FEAP}
\mathcal{F}_{\Omega}=\int_\Gamma f_{\Omega}(\vect{r})\,d\vect{r}=\int_\Gamma
\left\{f_{poly}(\vect{r}) + f_{ions}(\vect{r}) + f_{el}(\vect{r})\right\}d\vect{r}\\
\end{equation}
The stationary value of this action is identical to $\Omega$, as indicated by the
subscript. The integrand $f_{\Omega}$ is an ``action density'' which has been expressed as
a sum of polymeric ($f_{poly}$), ionic ($f_{ions}$) and electrostatic ($f_{el}$)
contributions.

To describe the polymer contribution we adopt the square gradient approximation (SGA)
\cite{CahnHilliard}. In this approximation the local free energy density of a polymer
solution is given by the reference free energy of the solution at position $\vect{r}$ plus
a square gradient term accounting for the nonlocal effects of concentration
fluctuations on chain entropies. We choose the Flory--Huggins expression \cite{Doi} on a cubic lattice as the
reference free energy. The polymer contribution to $f_{\Omega}$ is thus:
\begin{eqnarray}\label{a_poly}
f_{poly}(\vect{r})& = &\frac{T}{a^3}
\left[\frac{\phi}{N}\ln{\frac{\phi}{N}}+(1-\phi)\ln(1-\phi)+\chi\phi (1-\phi)
-\frac{\mu_p}{T}\phi\right]+\nonumber \\ & & \,+\frac{T}{36 a} \frac{(\nabla \phi)^2}{\phi}
\end{eqnarray}
where $T$ is the temperature of the solution, $a$ is the lattice parameter of a cubic lattice, $\phi$
is the monomer volume fraction defined on such a lattice, $N$ is the number of monomers in a chain, $\chi$ is the Flory
interaction parameter, and $\mu_p$ is the monomer chemical potential, which is fixed by the
reservoir. The coefficient of the
square gradient term derives from a  comparison of the small fluctuation limit of the SGA
with the large wavelength limit of the free energy expansion of a polymer solution in the
random phase approximation (RPA) \cite{DeGennesBook, CatesPagonabarraga,
  NOTE_prefactor_SG}.  
The ionic contribution, as in Poisson--Boltzmann (PB) theory \cite{ElecEffSoftMatt},  
comprises an entropic ``ideal gas'' term and a chemical potential term added to ensure 
ion density conservation:
\begin{equation}\label{a_ions}
f_{ions}(\vect{r})=T\sum_{i=+,-} n_i (\ln{n_i}-1)-\sum_{i=+,-} \mu_i n_i
\end{equation}
where $n_i$ is the number density of ionic species $i=+,-$, and $\mu_i$ and $T$ are
respectively the chemical potential of species $i=+,-$ and the temperature of the system,
fixed by the reservoir. As in standard PB theory \cite{ElecEffSoftMatt}, we consider a dilute solution of pointlike ions. We thus ignore all steric effects associated with finite ion size. This is a good assumption since we are interested in the qualitative behaviour of small ions residing in a polymer solution which may be reasonably concentrated (see \sect{ModelIIResults}), but is sufficiently far from a melt that any steric reduction of the ion entropy by the polymer can be safely ignored.

\begin{figure}[tbph]
\centering  
\includegraphics[width=0.4\linewidth]{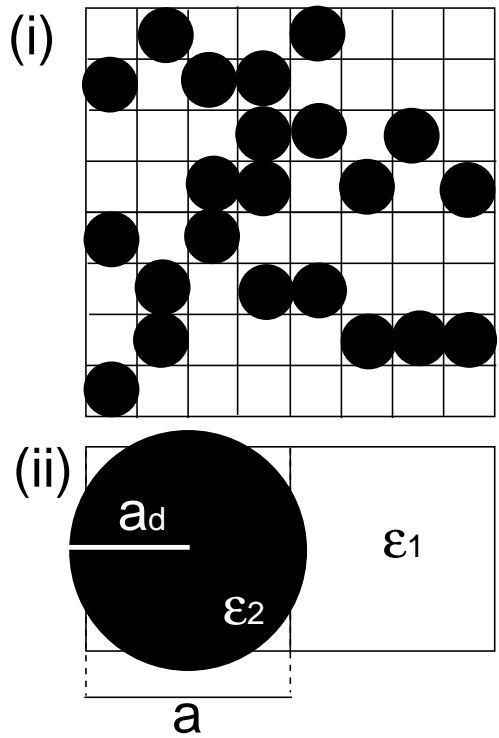} 
\caption{(i) The Flory--Huggins model on a cubic lattice, here represented by a two 
  dimensional square lattice. (ii) Closeup of two cells of the lattice. A 
dielectric sphere of permittivity
  $\epsilon_2$ and radius $a_d \neq a$ is associated with each occupied site. The surrounding
  solvent has the same permittivity $\epsilon_1$ of a pure electrolytic solution. The relation 
between $a_d$ and $a$ is discussed in the text.}\label{lattice_figure} 
\end{figure}

The electrostatic contribution to the free energy action is:
\begin{equation}\label{a_el}
f_{el}(\vect{r})=\sum_{i=+,-} n_i q_i V -\frac{1}{2}\epseff(\phi)(\nabla V)^2
\end{equation}
where $q_i=z_i e$ is the charge of species $i$ with valence $z_i$, and $V$ is the
electrostatic potential. \eqn{a_el} represents the variational formulation of
electrostatics \cite{Schwinger}: Poisson's equation follows from the variation $\delta
\mathcal{F}_{el}=\int_\Gamma f_{el}\,d\vect{r}=0$ and if its solutions are substituted in
\eq{a_el}, the electrostatic energy of the system follows \cite{Schwinger}. Throughout
these equations we have set $k_B = 1 = \epsilon_0$.

The dielectric nature of the polymer solution is accounted for via the effective
permittivity \epseff, which is a function of the monomer concentration 
$\phi(\vect{r})$ at $\vect{r}$. The polymer solution is thus treated as a dielectric
mixture of solvent with permittivity $\epsilon_1$ and ``dielectric monomers'' with
permittivity $\epsilon_2$. Since the mean field treatment of mixtures with spherical
inclusions is particularly simple, we assign a spherical
dielectric volume with characteristic radius $a_d$ to each monomer, as shown in
\fig{lattice_figure}. In general, we expect 
$a_d$ to differ from the cubic lattice length $a$ by an unknown factor $\gamma$ that depends
on the local chain geometry, so that
$a_d=a/\gamma$. 
We can now use the Maxwell--Garnett equation for the effective permittivity of
a mixture with spherical inclusions \cite{Sihvola}:  
\begin{equation}\label{MGphipoly}
\epseff(\phi)=\epsilon_{1} \left(1-\frac{3 K \alpha\phi}{1+K \alpha \phi}\right)
\end{equation}
Here $K\equiv(\epsilon_1-\epsilon_2)/(2\epsilon_1+\epsilon_2)$ is the Mossotti
factor. The coefficient $\alpha \equiv (4/3)\pi / \gamma^3$ accounts for the difference
between the volume of a cubic cell of the lattice and that of the effective dielectric sphere
each of which models, in its own fashion, a single monomer.

Effective medium approaches, such as that leading to \eqn{MGphipoly}, involve averaging
the electric and displacement fields over a region containing discrete inclusions. When
such averaged fields are used in place of the real ``microscopic'' fields, the medium can
be properly characterised by an effective permittivity. This will be a smooth function provided
the averaging volume is large enough to contain several inclusions \cite{LandauED, SipeBoyd}. Thus 
\eqn{MGphipoly} is strictly valid only if $R \gg a \phi^{-1/3}$, where $R$ 
is the radius of a spherical averaging volume, containing inclusions of size $a$ and
number concentration $\phi / a^3$. If the electric field itself is not mesoscopically uniform but 
varying on some
length scale $\lambda$ (in our case the Debye length), we also require
$\lambda \gg R$, so that the field which is being averaged does not change in 
magnitude over the averaging volume itself.

Aside from this, our model is subject to the same approximations as the 
Poisson--Boltzmann and Flory--Huggins mean field theories. In particular, the mean field
approximation implies that all fluctuation-induced effects are ignored and that the theory
is strictly inapplicable to dilute or semi-dilute polymer solutions in good solvents, where Flory--Huggins
fails; however we expect it to give a reasonable account of the global picture. 
On top of this, use of the square gradient approximation requires
externally induced concentration  variations (and therefore the inducing 
external fields) to be slowly varying. Another tacit assumption is that the direct
effect of ions on the solvent--monomer interactions expressed by the Flory $\chi$ parameter is
negligible; this might not be true in the presence of complexation between chain and ion but
should be adequate otherwise. 

\subsubsection{Model Equations}

The model's equations follow from the requirement that \eqn{FEAP} be stationary with
respect to variations in $\phi$, $n_i$ and $V$ (\appdx{Appdx}). With the additional simplifying
assumption that all the ions in solution are monovalent and with the substitution
$\phi=\psi^2$ for the polymer concentration variable, the equations read:

\begin{eqnarray}
\nabla\left[\epseffbar(\psi)\nabla \left(\frac{eV}{T}\right)\right]&=& \kappa^2\sinh\left(\frac{eV}{T}\right)\label{PBPolyMono}\\
\frac{a^2}{9}\nabla^2\psi&=&\label{PolyPolyEq0} \frac{3}{2}\frac{K\epsilon_1 \alpha a^3}
{T}(\nabla V)^2 \frac{\psi}{(1+K \alpha\psi^2)^2} +\\ & & +
\frac{\psi}{N}\ln{\left(\frac{\psi^2}{\psi^{r\,2}}\right)}- \psi \ln\left(
\frac{1-\psi^2}{1-\psi^{r\,2}} \right)- 2\chi(\psi^3-\psi^{r\,2}\psi)\nonumber
\end{eqnarray}
where $\epseffbar\equiv\epseff/\epsilon_1$ is given by the Maxwell--Garnett
\eqn{MGphipoly}, $\kappa\equiv \sqrt{2 n_s^r e^2 / \epsilon_1 T}$ is the inverse Debye
length for a monovalent salt of number density $n_s^r$, and the 
superscript $r$ indicates reservoir values of salt or polymer concentrations.

\eqn{PBPolyMono} is the Poisson--Boltzmann equation with a permittivity that depends on polymer
concentration and hence on position. \eqn{PolyPolyEq0} describes the concentration variations of a
dielectric polymer solution. Similar mean field descriptions of polyelectrolyte adsorption
\cite{BorukhovAndelman, NetzAndelman, ElecEffSoftMatt} and its effect on surface interactions
\cite{BorukhovAndelman, NetzAndelman} have been carried out. Like the latter, Equations
\eq{PBPolyMono} and  \eq{PolyPolyEq0} describe the concentration variations of a polymer
solution next to charged walls. However, since we are not modelling charged polymers, the
polymer profile described by \eq{PolyPolyEq0} is only electrostatically affected because
of the dielectric nature of the polymers (described by the first term on the right hand
side of \eq{PolyPolyEq0}, which derives from the Maxwell--Garnett relation). 
Conversely, in \eqn{PolyPolyEq0} the polymer enters only
through the modified permittivity.

The influence of confining charged surfaces is incorporated into boundary conditions on
Equations \eq{PBPolyMono} and \eq{PolyPolyEq0}. Assuming that the surfaces are parallel,
flat, homogeneous and infinite reduces our problem to one dimension. We shall denote the
position variable by $x$, with origin on the ``leftmost'' surface ($x=0$, as shown in
\fig{PolymerChain}). $D$ indicates the separation between the surfaces (the ``rightmost''
surface is located at $x=D$). For simplicity we have chosen nonadsorbing polymers, and also 
now choose
identical surfaces with fixed surface charge; these choices translate into the following boundary
conditions:
\begin{eqnarray}\label{BC1}
V^{\prime}(0) &=&  -\frac{\sigma}{\epsilon_1} = - V^{\prime}(D) \\   \psi(0) &=& 0 =
\psi(D)\label{BC2}
\end{eqnarray}
where the primes mean $d/dx$.  The fixed
surface charge condition was chosen in view of describing charged lamellar phases, whose
unknown surface charge density can be estimated from the area available to charged
surfactant groups. (It would also be possible to impose fixed surface potential, but
this is more complicated numerically and we do not attempt it here.)

The solution of Equations \eq{PBPolyMono} and \eq{PolyPolyEq0}, subject to Equations
\eq{BC1} and \eq{BC2}, allows evaluation of the force per unit area acting between
surfaces in the presence of polymer. This is the `net osmotic pressure' $\Pi^{net} = \Pi - \Pi^r$, defined
as the osmotic
pressure of ions and polymers on the midplane, $\Pi$, less that in the reservoir, $\Pi^r$. Note that on the midplane, and also in the reservoir, the local osmotic pressure equates to the normal ($xx$) component of
a stress tensor $\Sigma_{\alpha\beta}$ that elsewhere includes the 
Maxwell stress arising from electric fields. The Maxwell stress
vanishes in the reservoir, and on the midplane by symmetry, which is why the normal
force can be equated to the net osmotic pressure there. To find $\Pi$, we  
first show in \appdx{Appdx} that the stress component $\Sigma_{xx}$ is independent of $x$. 
Evaluating this (on the midplane
for convenience) as a function of plate separation $D$ gives $\Pi(D)$ and hence $\Pi^{net}(D)$. From
\eqn{PolyPiMono} of \appdx{Appdx}, evaluated at the midplane $x=D/2$, we finally obtain:

\begin{eqnarray}\label{PolyPiMono0}
\Pi^{net}(D)& = & 4 n^r_s T \sinh^2\left(\frac{eV|_{D/2}}{2T}\right) -
\frac{T}{a^3}\left[\frac{\phi|_{D/2}}{N}\ln{\frac{\phi|_{D/2}}{\phi^r}}
+\frac{(\phi^r-\phi|_{D/2})}{N}+\right.\\ & & \left. +
(1-\phi|_{D/2})\ln\left(\frac{1-\phi|_{D/2}}{1-\phi^r}\right) - \chi
(\phi^r-\phi|_{D/2})^2- (\phi^r-\phi|_{D/2})\right]\nonumber
\end{eqnarray}
\eqn{PolyPiMono0} includes both ionic and polymeric contributions to the net pressure: the first term on the
right hand side of  the equation and the term in square brackets, respectively. The
ionic contribution is always repulsive (as one would expect from a mean field
treatment of the electrostatics). The polymeric contribution, however, can become
attractive at surface separations which unfavourably confine the polymer (values of $D$
such that $\phi|_{D/2}<\phi^r$). We find below that the interplay of such opposing forces can lead to phase
separation, analogous to the liquid--gas transition of ordinary fluids. At mean field
level, this shows up as a characteristic $S$--shaped loop in the net osmotic pressure (like
that of the isotherms predicted by the Van der Waals equation of state \cite{LandauSP}). Our model 
can thereby be used to predict
the equilibria of lamellar phases, which are known to phase separate when polymer is
added to them (see \cite{KotzKosmella} and references therein).

\subsection{Mapping onto Lamellar Phases}

\begin{figure}[tbph]
\centering
\includegraphics[width=0.5\linewidth]{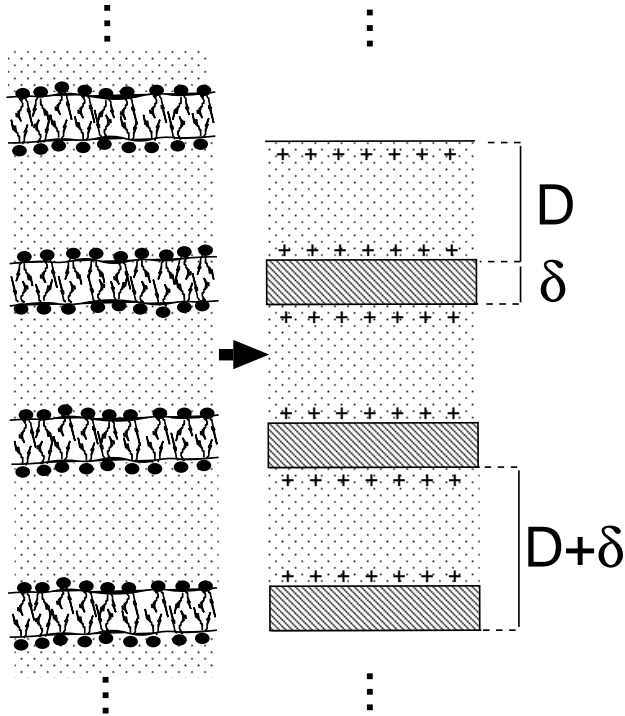}
\caption{A polymer-doped lamellar phase containing $N_b$ bilayers, 
  approximated by a periodic succession of units. Each unit
  comprises a polymer solution confined between charged parallel planes (as 
  described by our model) and a rigid rectangular slab (the bilayer).}\label{Mapping}
\end{figure}

A lamellar phase consists of a periodic one
dimensional crystal of repeating units. Each unit comprises a bilayer of width $\delta$
and an adjacent solvent layer of width $D$, so that the crystal's repeat distance is
$D+\delta$.
Consider a lamellar phase containing $N_b$ bilayers in contact with a polymer and ion
reservoir via a membrane impermeable to the bilayers. The solvent layers of such a
phase can be modelled as a  polymer solution confined between flat parallel surfaces; a
situation described by the model we have just built. This mapping, shown in \fig{Mapping},
is reasonable for lamellar phases with rigid bilayers (so that the surfaces are
approximately flat and $D$ is well defined). Further, since our model does not account for
other known forces between bilayers (dispersion, hydration, Helfrich etc.) the mapping is
strictly valid only when electrostatics and polymer-induced forces dominate.

Under these assumptions our model can predict the thermodynamic behaviour of a
lamellar phase, which is entirely determined by the free energy of the solvent
slab. 
Since pressure is an intensive thermodynamic variable,
the osmotic pressure of the lamellar phase is identical to \eqn{PolyPiMono0}; this is
the equation of state of the lamellar phase, and determines its phase
behaviour.

\section{Results: Parallel Surfaces \label{ModelIIResults}}

In what follows we give results from the numerical solution to our
model. \tab{phaseparameters} displays the ``baseline parameters" used in numerical
evaluations, all but one of which will be held fixed at the values shown while the one
remaining parameter is varied. The baseline values are loosely based on experimental systems
\cite{LigoureTheoryExp97,LigoureElec97,Ligoure2000,Javierre01,Linse1}:
typical charged bilayers (e.g. \rm{CPCl} or \rm{SDS} mixed alcohol) in an aqueous salt solution
of water--soluble polymers (\rm{PVP} or \rm{PEG}) at room temperature. However, particular
parameters were adjusted to have values which might enhance the effects of dielectric coupling,
creating on purpose a `worst case' scenario for the additive approximation \cite{NOTE_approx_sat}. 

\begin{table}[tbph]
\begin{center}
\begin{tabular}{l c c}
\hline 
\if{Elementary charge & $e$	 & $1.6\times 10^{-19}\,\rm{C}$  	    \\
Permittivity of vacuum & $\epsilon_0$ & $8.85\times 10^{-12}\,\rm{F\,m^{-1}}$ \\
Boltzmann's constant & $k_B$	 & $1.38\times 10^{-23}\,\rm{J\,K^{-1}}$ \\ 
}\fi
Temperature  & $T$	& $298\ut{K}$\\ 
Permittivity of ionic solution & $\epsilon_1$ & $78.5$ \\ 
Permittivity of polymer & $\epsilon_2$ & $2$ \\ 
Dielectric factor & $\gamma$ & $1.5$ \\   	        	       
Polymer lattice length & $a$ & $10\ut{\AA}$\\  
Flory Parameter & $\chi$ & $0.495$ \\ 
Number of lattice units per chain & $N$ & $2000$\\ 
Surface charge density & $\sigma$& $0.1\ut{e\,nm^{-2}}$ \\
\if{(Area per surface charge) & ($\Sigma$) & ($1000\ut{\AA^2\,e^{-1}}$)\\
}\fi
Reservoir salt concentration & $c_s^r$ &$0.02\ut{M}$\\
\if{(Debye length) & ($\lambda$) & ($21.5\ut{\AA}$)\\
\fi
Reservoir monomer volume fraction  & $\phi^r$ & $0.3$\\
\hline
\end{tabular}
\caption{Values of the parameters used in the numerical evaluation of the model equations
  and the subsequent determination of phase diagrams. The Debye length in the reservoir
  corresponding to the salt concentration shown above is $\lambda=21.5\ut{\AA}$.} \label{phaseparameters} 
\end{center}
\end{table}

To maximise dielectric contrast, a value of $2$ (the permittivity of hydrocarbon oils) is
chosen for the polymer permittivity $\epsilon_2$. 
Similarly, the $\chi$ parameter is set near the {\it theta} point,
enhancing the susceptibility of the polymer to external fields.
Our choice of $\gamma$, the size difference between the lattice length and the radius of the
dielectric sphere, is likewise made to enhance coupling effects. Physically, the dielectric
size of a lattice monomer depends on how many ``oily'' hydrocarbon groups are contained in
the backbone or side groups of every chemical monomer. How much bigger the dielectric
volume occupied by the polymer is with respect to the steric volume is hard to estimate
precisely.
Our choice of $\gamma=1.5$ entails that the Flory--Huggins (``entropic'') volume available to a lattice
monomer, $V_{FH}$, occupies about $80\%$ of the polarisable volume, $V_d$ ($V_{FH}/V_d
\simeq \gamma^3/4 \approx 0.8$). 
The lattice length $a$ is modelled on the water soluble polymer PVP. As shown in
\cite{Cosgrove}, $a$ can be related to the chemical monomer size, $l$, of the polymer of
interest and the polymer molecular weight, $M_w$, to the number $N$ of lattice units per
chain. On a cubic lattice, one finds $a=10\ut{\AA}$ and $N=2000$ for PVP with $l\simeq 30
\ut{\AA}$ and $M_w\simeq500000$. (Note that the $N$ is not an important control parameter
in the parameter regime of interest to us.)
Finally the reservoir monomer concentration was chosen at a reasonably high value of
$\phi^r=0.3$ so that the concentration of polymer would not be too small, swamping out any
dielectric coupling effects. 

Given these ``worst-case" choices, the uncoupled theory gives a surprisingly good
approximation to the full one in most cases. To a significant extent, this justifies the
assumption of additivity, tacitly made by some authors \cite{Russel}, and inconsistently
justified by others \cite{Sear1, Ligoure2000}. However, we are not aware of any simple
order--of--magnitude arguments that can explain this without pursuing the more detailed
calculations presented here. 
 
\subsection{Potential and Polymer Concentration Profiles \label{PhysPredModIIEqnsSect}}

\begin{figure}[tbph]
\centering
\centerline{\includegraphics[width=0.5\linewidth]{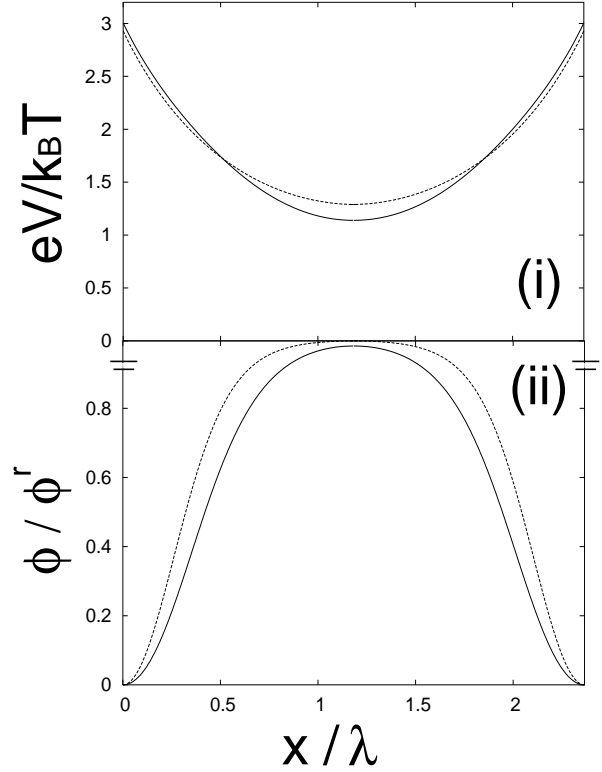}}
\caption{Plots of the numerical solutions of Equations \eq{PBPolyMono} and
  \eq{PolyPolyEq0} subject to the boundary conditions \eq{BC1} and \eq{BC2}, for a surface
  separation is $D \simeq 2.37 \lambda$, and using the parameters of
  \tab{phaseparameters}. The results display 
  both coupled (solid line) and uncoupled 
  (dotted line) solutions for (i) the dimensionless
  electrostatic potential, $eV/k_BT$, and (ii) the monomer concentration rescaled by the
  reservoir value, $\phi/ \phi^r$, as functions of the dimensionless position
  between surfaces, $x/ \lambda$.}\label{PolySolns}
\end{figure}

\fig{PolySolns} displays the dimensionless electrostatic potential $W\equiv e V /T$ and
monomer concentration profile $\Phi\equiv \phi/\phi^r$ as a function of dimensionless
position $X\equiv x/\lambda$, for surfaces separated by $D / \lambda\simeq 2.37$ (recall
$\lambda$ is the Debye length defined in the reservoir). The solutions to the full model
(coupled equations) were obtained using an adaptation of the shooting method
\cite{NumericalRecipes} to solve the 
Equations \eq{PBPolyMono} and \eq{PolyPolyEq0} subject to the
boundary conditions \eq{BC1} and \eq{BC2}. 
For comparison, the solutions found in the additive approximation (uncoupled
equations) are also presented.

Qualitatively, the solutions of the two descriptions are similar. The ``coupled"
electrostatic potential displays the characteristic symmetric shape of the solutions to
the standard Poisson--Boltzmann equation (uncoupled case); similarly, the monomer
concentration profile compares well with the predictions of a mean field description of
nonadsorbing polymer 
chains confined in a slit \cite{Teraoka}. The characteristic depletion of monomers from
between the surfaces can be observed in both cases. It is well known that the pressure
imbalance created by such depletion can cause the surfaces to attract, if the polymer
solution is sufficiently squashed that it escapes from between the plates, reducing
the osmotic pressure there.

The results, however, also highlight the interesting features which follow from a full
account of the polymers' dielectric properties. First, the electrostatic potential
is reduced around the midplane (where the polymer monomers are more concentrated) with
respect to the uncoupled solutions. The reduction is due to the additional electrostatic
screening provided by the monomers because of their polarisation. Second, the monomer
concentration profile displays an increased depletion from the surfaces. This is also a
consequence of polarisation and results from the electrostatic energy penalty of placing
dielectric monomers in the surface fields.

\subsection{Osmotic Pressure Between Parallel Surfaces \label{ModIIOsmPress}}

\begin{figure}[tbph]
\centering
\centerline{\includegraphics[width=0.8\linewidth]{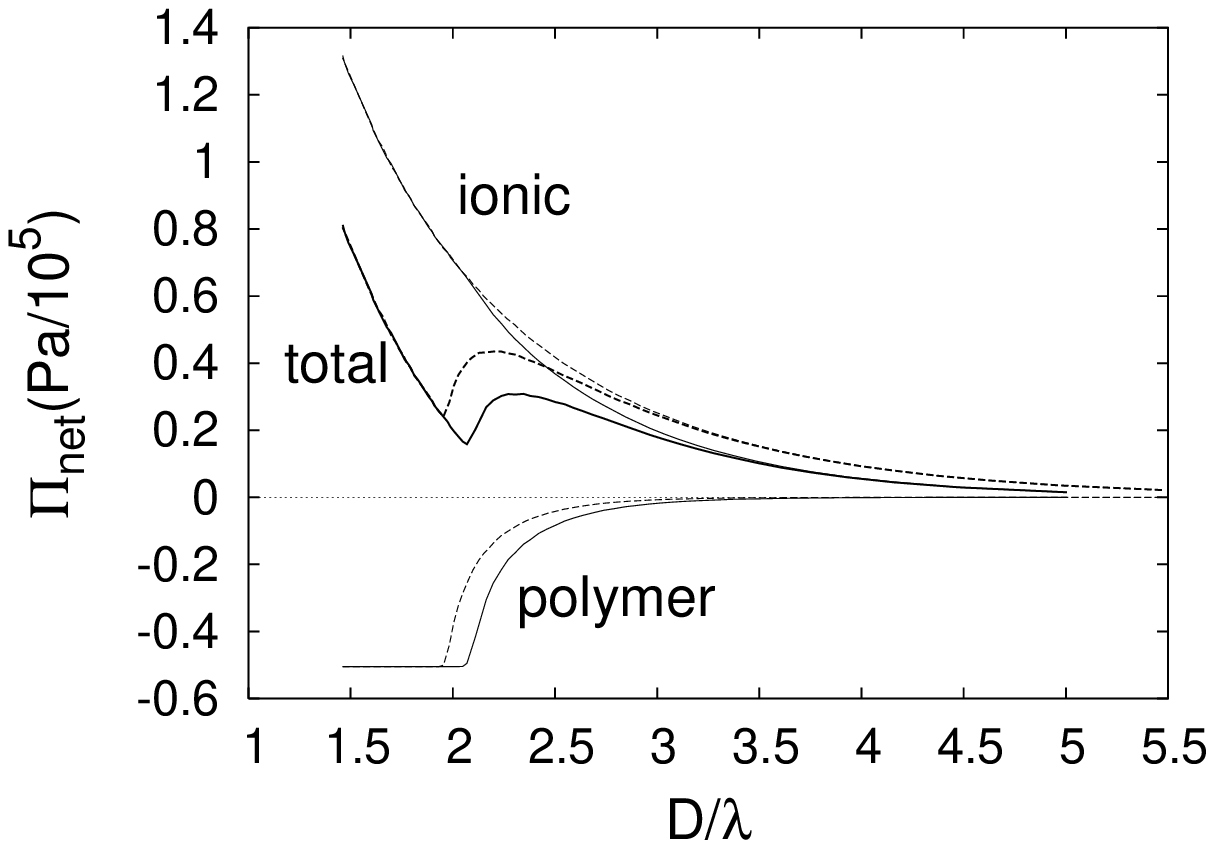}}
\caption{Net osmotic pressure as a function of 
  $D/\lambda$, together with its ionic and polymeric
  contributions. The plots were obtained from \eqn{PolyPiMono0} by substituting the midplane 
  values of $V$ and $\phi$ from the solutions of \eq{PBPolyMono} and
  \eq{PolyPolyEq0} for various $D$. The  coupled (solid
  lines) and uncoupled (dashed lines) cases are shown. Parameters as
  in \tab{phaseparameters}.}\label{SigOsmoticPress}  
\end{figure}

\fig{SigOsmoticPress} displays the predicted variation (coupled and uncoupled cases) of total osmotic pressure between
surfaces of dimensionless surface separation $D/\lambda$. Both descriptions predict 
a van der Waals loop. However, in the coupled picture the
loop is translated to lower pressures and slightly larger separations with respect to the
uncoupled, additive case. In addition, while at very small separations coupled and uncoupled
profiles coincide, at large separations dielectric coupling lowers the osmotic pressure
curve below the additive prediction.

To understand these differences the ionic and polymeric contributions to the midplane
pressure are also shown in \fig{SigOsmoticPress}. For both models, the
ionic contribution is always repulsive, whilst the polymeric contribution is
attractive for separations below those at which the polymer starts to be expelled from
between the plates. However, for distances large enough that no significant amount of
polymer has been expelled, the ionic contribution to the coupled pressure profile
decays more rapidly than the uncoupled prediction.  This stems from the
more efficient screening of the electrostatics in the presence of dielectric
monomers. Another effect of the dielectric coupling is that the
polymer contribution becomes attractive at larger $D/ \lambda$ than the uncoupled one. This
represents an ``electrostatically enhanced'' depletion of polymer
from the surfaces as the dielectric monomers are expelled from the high--field regions
near the confining plates.

We can thus summarize the overall differences between the net pressure profiles: at
large distances the shorter range of the coupled electrostatic repulsion lowers the
pressure profile with respect to the uncoupled prediction; the larger range of the polymer
expulsion then promotes the occurrence of the van der Waals loop; finally, at small
enough separations the polymer is fully expelled, the dielectric coupling is
removed, and coupled and uncoupled predictions coincide.

\section{Results: Phase Diagrams \label{ResPhasDiags}}

\subsection{Expected Phase Behaviour}

As mentioned in the Introduction, the addition of polymer to charged lamellar phases has
experimentally been observed to induce phase separation. X--ray scattering studies on
these systems have mainly evidenced two kinds of separation \cite{Ficheux95,
  LigoureTheoryExp97}: $\rm L_\alpha 
L_\alpha$ coexistence between two lamellar phases with  
different spacings, and $\rm L_\alpha L$ coexistence between a
single lamellar phase and an isotropic solution of polymer (with trace surfactant). 
The occurence of phase separation and its modality depend, if all other
parameters are held fixed, on the relative composition of the prepared mixture. This has
been usefully mapped on density--density phase diagrams plotting the polymer content of
the sample against its surfactant composition (e.g. \cite{Ficheux95, Javierre01}).

\subsection{Predicting Phase Coexistence}

\begin{figure}[tbph]
\centering
\includegraphics[width=0.5\linewidth]{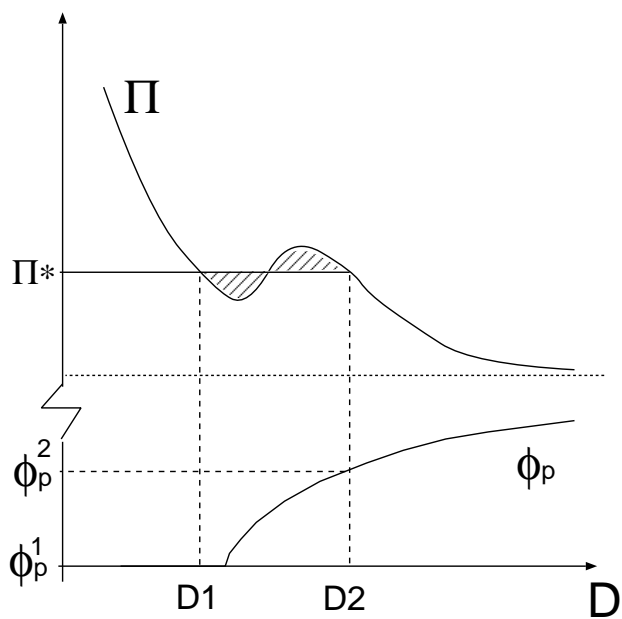}
\caption{The Maxwell construction for the osmotic pressure 
\eq{PolyPiMono0} of a polymer doped lamellar phase, 
predicting $\rm L_\alpha L_\alpha$ coexistence at spacings $D_1,D_2$. Once these spacings are
identified, the corresponding polymer content of coexisting phases can be obtained from a
graph of \eqn{PolyContent}, as shown.}\label{maxwell_fig}    
\end{figure}

We can explain the phase behaviour just described using our model's equation of
state, \eqn{PolyPiMono0}. Recall that in mean field theories like ours, phase coexistence 
shows up as a Van der Waals loop in plots of the net osmotic pressure as a
function of bilayer separation $D$. To deal with this we can deploy 
the Maxwell construction \cite{LandauSP}, finding a horizontal line which cuts the 
van der Waals loop into two
regions of equal area (Fig. \ref{maxwell_fig}). The horizontality of the line represents the equality of  
pressures for coexisting phases of different spacings $D$, 
whereas the equal areas
represent the equality of chemical potentials
\cite{LandauSP}. The coexistences found in this way connect two lamellar phases
of different layer spacings ($\rm L_\alpha L_\alpha$ coexistence). This type of
coexistence only occurs for a positive pressure: referring to \fig{SigOsmoticPress}, the
polymer-induced 
attraction causes phase separation, but cannot overcome the electrostatic repulsion
between bilayers. 

\begin{figure}[tbph]
\centering
\includegraphics[width=0.5\linewidth]{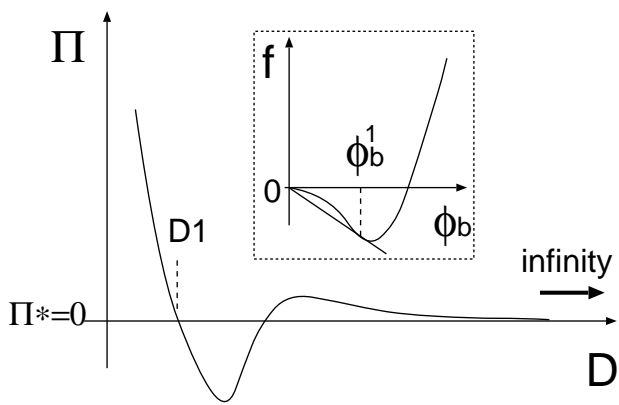}
\caption{The special Maxwell construction for $\rm L_\alpha L$
  coexistence. The isotropic fluid is represented by a putative lamellar phase with
  $D=\infty$ and hence zero osmotic pressure. In reality an
  exponentially small amount of surfactants coexisting with 
  the lamellar phase resides in the reservoir. The inset shows the equivalent condition of
  virtual tangency, when the common tangent construction is used to determine phase
  equilibria from the free energy density $f$ as a function of volume fraction $\phi_b$;
  see \protect\cite{BrooksMEC1}.}\label{virtual}    
\end{figure}

However, when the polymer contribution is large enough, the pressure can
become negative and a special type of Van der Waals loop occurs across the zero-pressure axis
(Fig. \ref{virtual}). This amounts to a coexistence 
between a lamellar phase of finite $D$, found where the pressure profile crosses
the zero axis, and a lamellar phase of infinite $D$ (identical to the reservoir) which is
represented
asymptotically by the pressure tending to zero at infinite $D$. 
Hence this modified construction allows the
prediction of $\rm L_\alpha L$ coexistence. However, the resulting 
areas are not equal in general; this is because the chemical potential of the surfactant
is undefined in the bilayer-free state of infinite $D$.  
Recalling that the Maxwell construction is
equivalent to the common-tangent construction on the Helmholtz free energy, the
failure to equate chemical potentials under these conditions corresponds 
to the so-called `virtual tangency'
condition 
in which a phase at finite surfactant density connects with one at zero density \cite{BrooksMEC1}. 
A more accurate treatment would
allow for the finite molecular or micellar solubility of surfactant in water, 
giving a slightly more elaborate calculation (in which the
slope of the free energy is rapidly varying at very low surfactant concentration) 
but an almost identical result for the coexistence properties.

Thus, using the reservoir monomer
concentration to control the polymer content of our lamellar
phase, \eqn{PolyPiMono0} allows us to determine the equilibrium spacings $D$.  The relation between
bilayer volume fractions $\phi_b$ and separations $D$ follows from simple geometry:
\begin{equation}\label{BilVolFrac}
\phi_b = \frac{\delta}{D+\delta}
\end{equation}
Where we recall $\delta$ is the bilayer thickness. Similarly, the bulk polymer
volume fraction $\phi_p$ within a lamellar phase is found by integrating the polymer profile
$\phi_p(x)\equiv\psi^2(x)$ (from the solution of \eqn{PolyPolyEq0}) for a given
$D$, normalising by the full volume now including the bilayers themselves:
\begin{equation}\label{PolyContent}
\phi_p= \frac{1}{D+\delta} \int_0^D \phi_p(x)dx
\end{equation}

\subsection{Semi-Grand Ensemble for Salt}
\label{SemiGrand}
At fixed chemical potential for salt, the preceding constructions allow 
us to find all states
of coexistence and the polymer and surfactant densities in them; below we create
phase diagrams essentially by `joining the dots' when these coexistences are plotted on the
$(\phi_b,\phi_p)$ plane. In most experiments, however, the experimenter fixes the total amount of
salt (as well as that of bilayer and polymer) in the system.
To deal with this is possible in principle, by a similar integration: the ion concentration
obeys \eq{PolyBoltzIons0}, so that defining the total salt concentration as $n_s\equiv n_+$, we
have:
\begin{equation}\label{SaltContent}
n_s= \frac{1}{D+\delta} \int_0^D n_i(x)dx =\frac{n_s^r}{D+\delta}\int_0^D \exp{-eV(x)/T}dx
\end{equation}
\eqn{SaltContent} expresses the Donnan equilibrium for a lamellar phase \cite{ZembDubois,
  Donnan}: since $V(x)$ is always positive for bilayers of finite separation, the total
  amount of salt $n_s$ within a lamellar phase is smaller than in the the reservoir. This
  salt expulsion is more 
efficient, the closer together the bilayers, so that coexisting phases with different
periods will contain different amounts of salt.

However, this additional calculation
represents a major numerical
complication. We have decided to neglect this, and
thus work in a semi-grand ensemble with respect to salt. 
Experimentally our phase diagrams are those of a lamellar
system in which a fixed volume of solvent, and fixed amounts of bilayer and polymer, are
in contact with a salt reservoir
through a dialysis membrane. This type of experiment is perfectly possible \cite{ZembDubois}, but is not
typical in studying polymer-doped lamellar phases. 
How the use of the salt reservoir affects our results will be discussed
in \sect{RelRealSystSect}.

\subsection{Effect of Bilayer Surface Charge Density \label{EffSurfChargDensSect}}

With these considerations in mind, we now present phase diagrams from our model by mapping
pairs of coexisting phase points. The error bars on the phase points result from the
uncertainties associated with the graphical method by which coexistences were found, or,
when this was very accurate, from the intrinsic accuracy of the numerics. The phase
boundaries themselves have been drawn as guides for the eye and do not represent
numerical fits to our data.

\begin{figure}[tbph]
\centering
\includegraphics[width=0.5\linewidth]{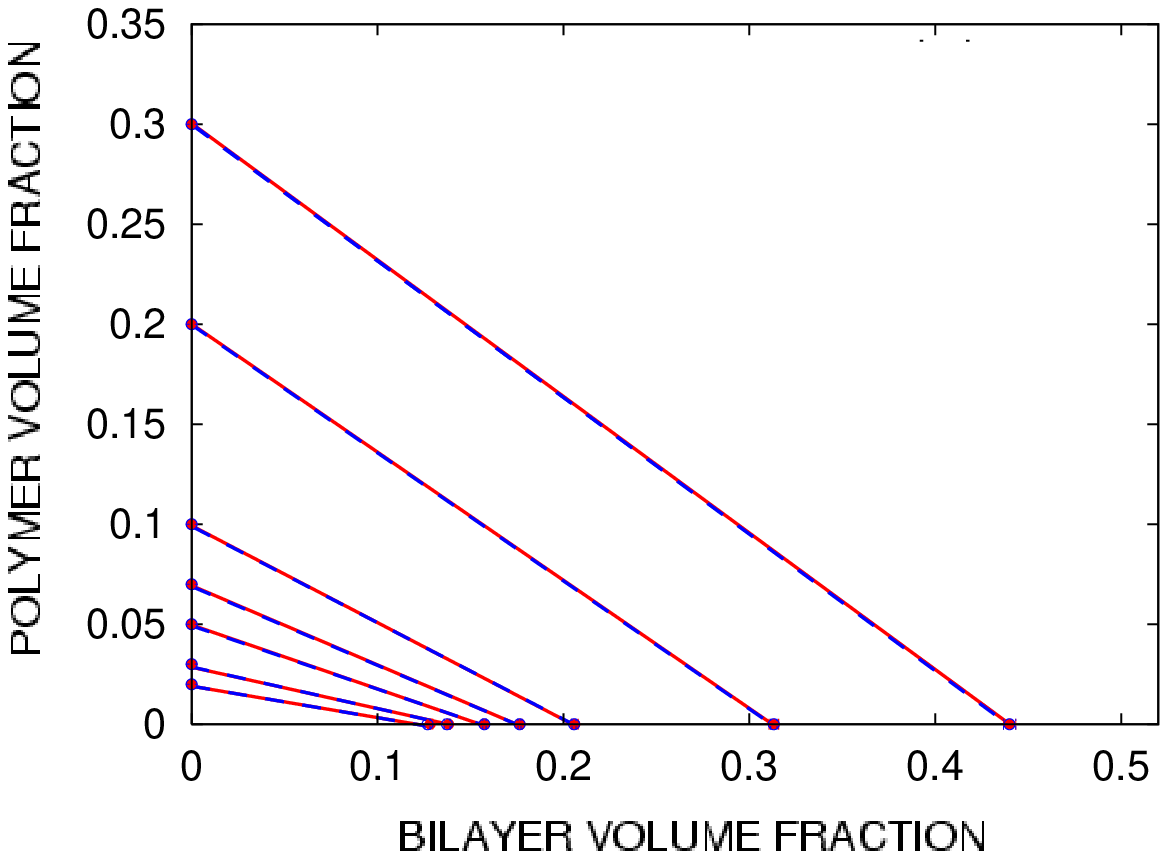}\\
\includegraphics[width=0.5\linewidth]{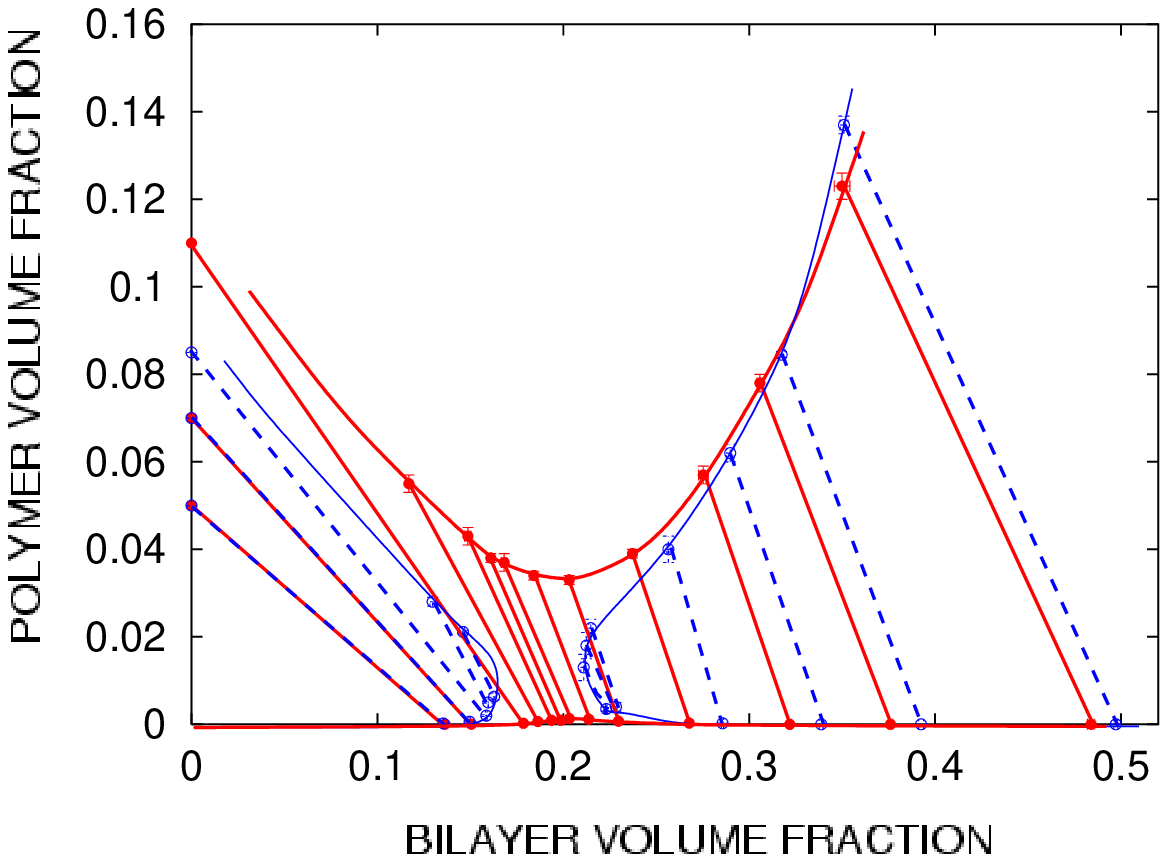}\\
\includegraphics[width=0.5\linewidth]{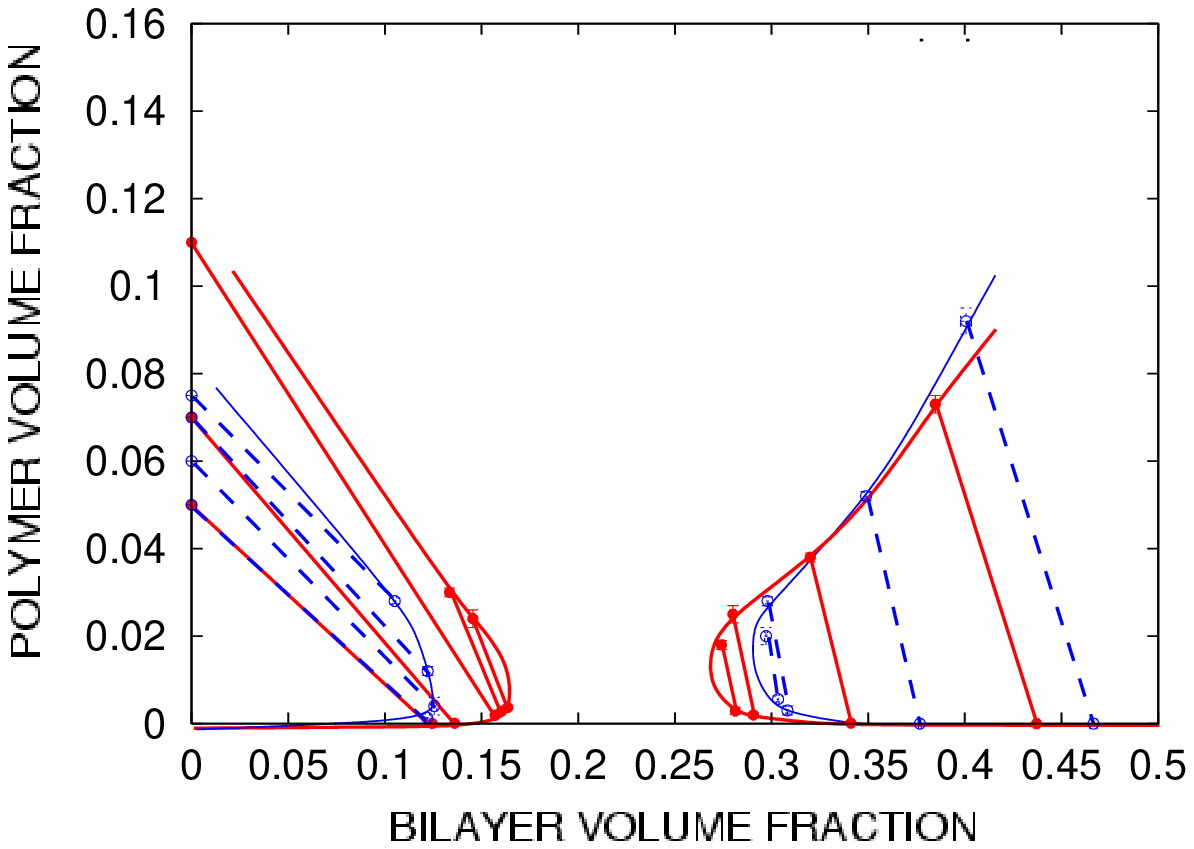}
\caption{Effect of increasing surface charge density, $\sigma$, on the phase behaviour of a polymer-doped
  lamellar phase in contact with a salt reservoir. Top to bottom: $\sigma=0.05
  \ut{e\,nm^{-2}}$, $\sigma=0.1 \ut{e\,nm^{-2}}$ and $\sigma=0.2
  \ut{e\,nm^{-2}}$. The coordinates of coexisting phases were found using the graphical
  method discussed in the text. Coupled ({\color{red}$\bullet$})  
  and uncoupled ({\color{blue}$\circ$}) points, with associated error bars, mark these
  coordinates and have been connected by tielines (full and broken
  respectively). Approximate binodals have also been fitted through the points to
  highlight the shape of the miscibility gaps. Note that the vertical scale on the top
  diagram extends further than that of the other two.} 
\label{sigma_study_fig}
\end{figure}

We study first the effect of changing the surface charge density,
$\sigma$. Density--density phase diagrams, with polymer and bilayer volume fractions as
composition variables, are shown in \fig{sigma_study_fig} for doped lamellar phases with
$\sigma=0.05, 0.1$ and $0.2 \ut{e\,nm^{-2}}$ (top, middle and bottom panels respectively). The 
middle phase diagram, calculated using the same parameters as
\tab{phaseparameters}, provides a reference for studying the influence of parameter
variation on phase behaviour. All other phase diagrams are calculated by changing the
parameter of interest above and below its reference value. 

Results are shown for both coupled and uncoupled equations. In both cases, an
increase in $\sigma$ changes the phase 
diagram topology from a large area of $\rm L_\alpha L$ coexistence
(top panel), through an intermediate  
``pinch off'' region (middle panel, coupled case), to two miscibility gaps, separated by
a single-phase $\rm L_\alpha$ region (middle panel, uncoupled; bottom panel). The $\rm L_\alpha L$ coexistence in the ``bilayer dilute'' region of the phase
diagram can be between the reservoir (isotropic phase) and polymer-free or
polymer-loaded lamellar phases. At higher bilayer volume fractions, only $\rm L_\alpha
L_\alpha$ coexistence between polymer-loaded lamellar phases is possible. Where the
density difference between coexisting $\rm L_\alpha$ phases vanishes, critical points
arise which close off the miscibility gaps; when this doesn't happen the two phases are
joined by a bottleneck of $\rm L_\alpha L_\alpha$ coexistence between loaded phases. At
very high bilayer volume fractions, $\rm L_\alpha L_\alpha$ coexistence is between a
polymer-loaded lamellar phase and one so concentrated as to contain no polymer.

The dielectric coupling, as could be expected, makes no difference at
very low surface charge densities (top panel of \fig{sigma_study_fig}).
At intermediate $\sigma$, however, (middle panel) the topology of the phase
diagram is coupling-dependent: in the coupled case, a characteristic
neck joins two miscibility gaps, whilst in the uncoupled
prediction the two gaps are already separate.  
At high $\sigma$ (bottom panel) the same
qualitative features are predicted for both approaches, but positions of phase
boundaries and critical points are visibly affected. In particular, the coupled model
predicts a dilute miscibility gap which is larger than in the uncoupled case.

The observed phase behaviour results from the interplay of polymer-induced attraction and
electrostatic repulsion between the bilayers. The general features of this are
not drastically modified the coupling. When $\sigma$ is small, the polymer physics dominates the
phase behaviour and drives $\rm L_\alpha L$ phase separation against a weak electrostatic
repulsion, for polymer and bilayer concentrations which span a large region of
the phase diagram. For higher values of $\sigma$ the polymer
attraction only dominate the electrostatic repulsion where this has decayed to
sufficiently small values. Alternatively, $\rm L_\alpha L_\alpha$ phase separation can occur
for concentrated bilayers, which confine the polymer solution and cause an attraction
sufficient to partially compensate the electrostatic repulsion.

It was shown in \sect{ModIIOsmPress} that including the dielectric coupling of the
polymers affects both electrostatic repulsion, whose range is decreased, and
the polymer-induced attraction, whose range is increased. For doped phases dilute in
both bilayer and polymer, these changes clearly broaden the extent of
the regions of phase separation (the pressure is more attractive over a greater range of
intralamellar separations because of coupling). Thus, the miscibility gaps
predicted by a model where electrostatics is coupled to polymer physics will be larger
than in an uncoupled description. At high polymer and bilayer volume fractions, however,  
the range of the attraction and of the repulsion are different, and
the coupled miscibility gap becomes smaller than the uncoupled prediction. This effect is
discussed in more detail \sect{EffectOfF}. 

\subsection{Effect of Reservoir Salt Content \label{EffSaltCont}}

\begin{figure}[tbph]
\centering
\includegraphics[width=0.5\linewidth]{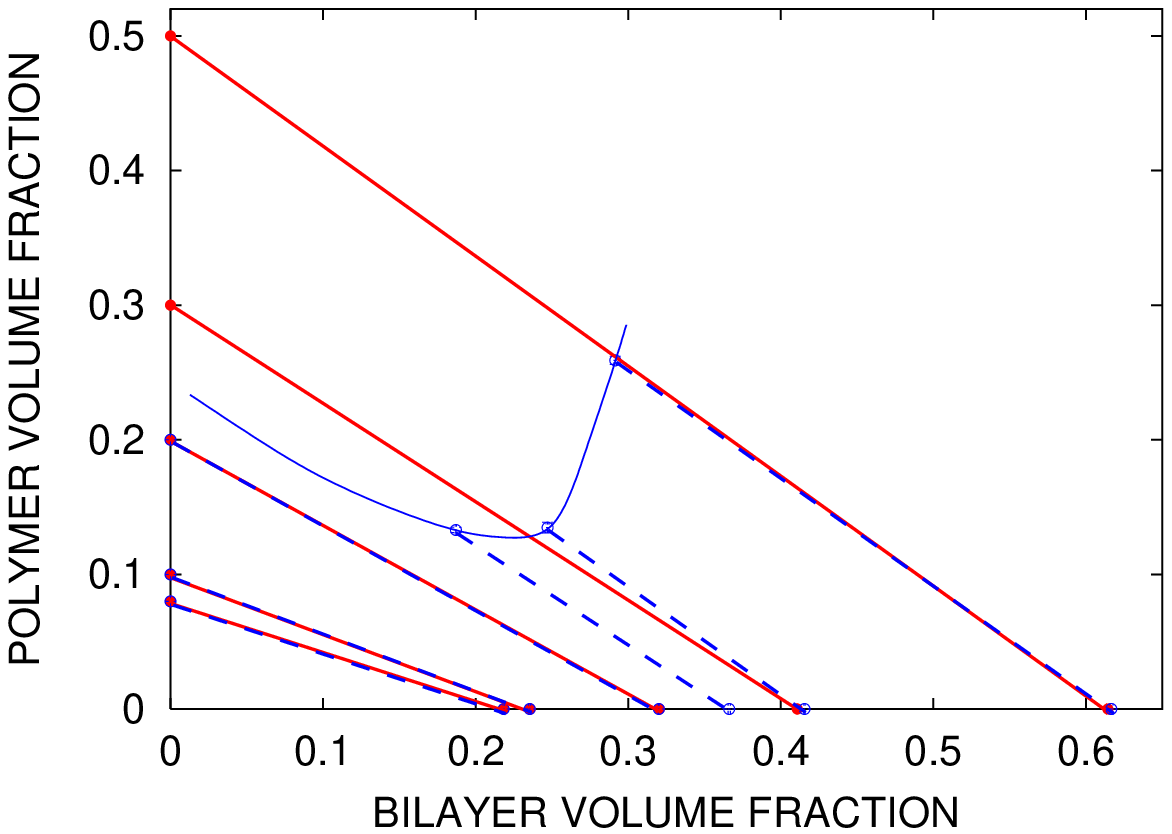}\\
\includegraphics[width=0.5\linewidth]{PaperFigures/PhaseDiagrams/s01c002phi-phi_phasediagtielined_PAPER}\\
\includegraphics[width=0.5\linewidth]{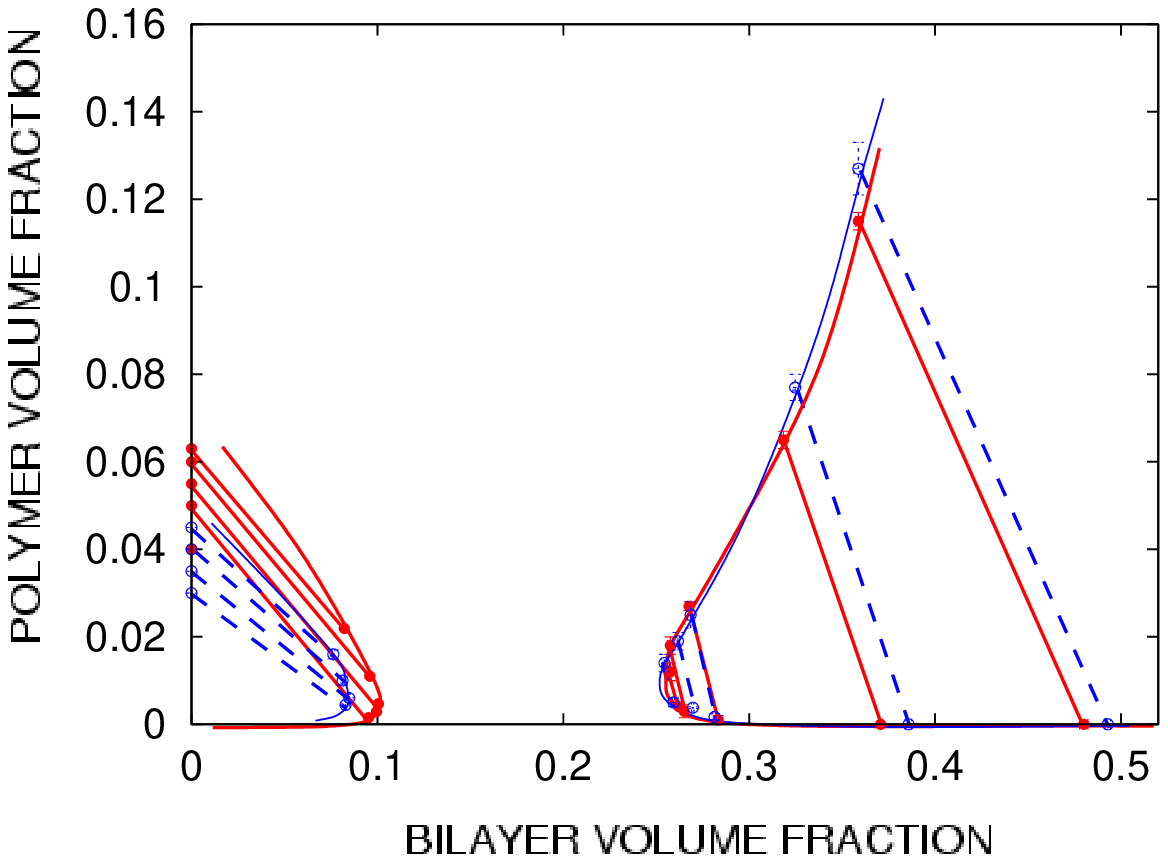}
\caption{Effect of decreasing the reservoir salt concentration, $c^r_s$, on the phase
  behaviour of a polymer--doped lamellar phase. Top to bottom: $c^r_s=0.047\ut{M}$,
  $c^r_s=0.02\ut{M}$, and $c^r_s=0.01\ut{M}$. 
Methods and notation as in Figure \protect{\ref{sigma_study_fig}}.
\if{
The coordinates of coexisting phases, were
  found using the graphical method discussed in the text. Coupled ({\color{red}$\bullet$})   
  and uncoupled ({\color{blue}$\circ$}) points, with associated error bars, mark these
  coordinates and have been connected by tielines (full and broken
  respectively). Approximate binodals have also been fitted through the points to
  highlight the shape of the miscibility gaps. }\fi
Note that the vertical scale on the top
  diagram extends further than that of the following two.}\label{salt_study_fig}
\end{figure}

Next we vary the salt concentration in the reservoir, to probe the
effect of the screening of electrostatic interactions on the phase
behaviour. \fig{salt_study_fig} shows the density--density phase 
diagrams when the reservoir contains $c_s^r=0.047 \rm M$, $0.02 \rm M$ and $0.01 \rm M$ of
salt (top, middle and bottom panels respectively; the middle panel is the same as in
the previous figure).

The effect of increasing the screening length $\lambda$ (decreasing salt) on phase
diagram topology is similar to that of increasing $\sigma$.   This is
no surprise, since reducing the salt level increases the range of the electrostatic
repulsion. The same general features arise: a large
$\rm L_\alpha L$ region (top panel, coupled); its pinch--off, which is delayed by 
coupling terms 
(top panel, uncoupled; middle panel, coupled); and finally two opposing miscibility gaps (middle panel,
uncoupled; bottom panel).
The relative extent of the phase regions and their changes in shape (tieline length
and tilt) evolves differently with $c_s^r$ than with $\sigma$, however. For example, the
miscibility gap in the dilute corner of the diagram shrinks in height as
$c_s^r$ decreases (middle to bottom panel), in contrast to Fig. \ref{sigma_study_fig}.

Just as for changes in $\sigma$, the dielectric coupling generally extends the
coexistence regions for a given $c_s^r$ compared to an uncoupled
description. In the top panel, $\rm  L_\alpha L$ coexistence is ubiquitous in the coupled predictions, 
whereas the uncoupled case shows $\rm L_\alpha L_\alpha$ coexistence 
beyond moderate bilayer concentrations. For $c_s^r=0.01 \rm
M$ (bottom panel), in the concentrated miscibility gap, almost no
differences between the two models are discernible (given the errors) 
in the phase boundary. The dilute coexistence region varies more
markedly, and is larger in the coupled case.

The basic mechanism of phase separation, and the effect of coupling on this, was
already described in the context of varying $\sigma$. Quantitative differences do
arise because here we are now changing the range, as
opposed to the magnitude, of the electrostatic repulsion. Dielectric coupling
modifies electrostatic screening somewhat, and also alters the polymeric interaction
by excluding monomers from regions of high field; these effects favour phase
separation as stated previously.

\subsection{Effect of the Solution Flory Parameter}

\begin{figure}[tbph]
\centering
\includegraphics[width=0.5\linewidth]{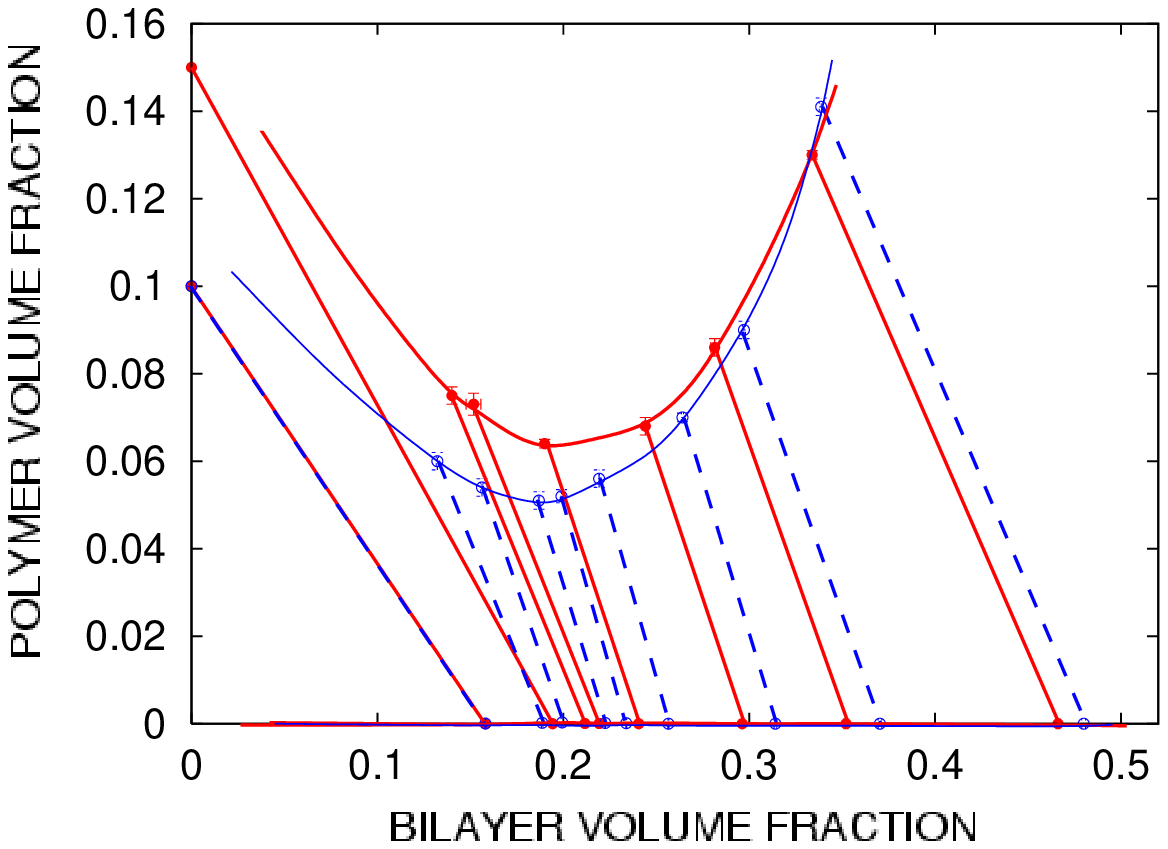}\\
\includegraphics[width=0.5\linewidth]{PaperFigures/PhaseDiagrams/s01c002phi-phi_phasediagtielined_PAPER}\\
\includegraphics[width=0.5\linewidth]{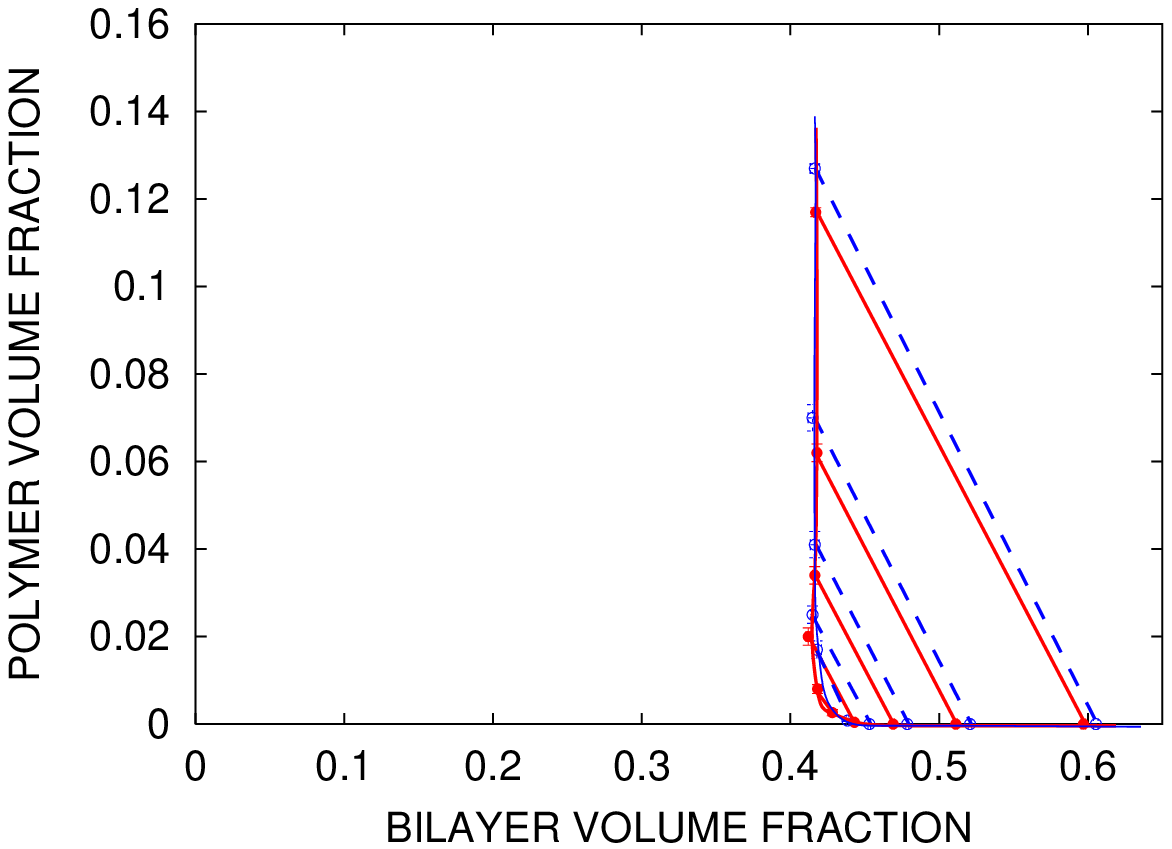}
\caption{Effect of decreasing the $\chi$--parameter on the phase behaviour of a
  polymer--doped lamellar phase in contact with a salt reservoir. Top to bottom:
  $\chi=0.515$, $\chi=0.495$, and $\chi=0.3$. Method and symbols as in Fig. \protect{\ref{sigma_study_fig}}
\if{The coordinates of coexisting phases, were
  found using the graphical method discussed in the text. Coupled ({\color{red}$\bullet$})   
  and uncoupled ({\color{blue}$\circ$}) points, with associated error bars, mark these
  coordinates and have been connected by tielines (full and broken
  respectively). Approximate binodals have also been fitted through the points to
  highlight the shape of the miscibility gaps.}\fi Note that the horizontal scale of the bottom
  diagram extends further than that of the preceding two.}\label{chi_study_fig}
\end{figure}

We now consider three different hypothetical polymers for which water is, respectively a poor solvent
with $\chi=0.515$ (\fig{chi_study_fig}, top panel); a near {\it theta} solvent with $\chi=0.495$
(middle panel: the ``reference'' system, as in previous figures); and a good solvent with $\chi=0.3$
(bottom panel). Decreasing $\chi$ in the range shown has broadly the same effect of an increase in
$\sigma$ or a decrease in $c_s^r$: merged $\rm L_\alpha L$ and $\rm L_\alpha L_\alpha$
regions (top panel) shrink  (middle panel, coupled) and split into
separate dilute and concentrated miscibility gaps (middle panel, uncoupled). Finally, the
dilute miscibility gap disappears and only the concentrated gap is left (bottom panel). 
The general features of this phase evolution arise because a variation in the
$\chi$ parameter affects both the strength and range of the polymer
interactions. Thus for poor and for {\it theta} solvents, the polymer solution can sense its
confinement at large separations, so that bound lamellar phases are found in both dilute and
concentrated regions. For a good solvent the range of the
polymer-induced attraction is too short to cause phase separation in dilute samples so this is only
effective at high concentration $\phi_b$.

The differences between coupled and uncoupled approaches are greater for the higher values
of $\chi=0.515,0.495$. The top and middle panels of \fig{chi_study_fig} show how the differences
are more pronounced on the bilayer-dilute side of the phase diagram, where once again the
coupling increases the extent of the $\rm L_\alpha L$ region with respect to the uncoupled
case. However, for good solvents this miscibility gap is absent; and even for the miscibility gap at
high concentration, coupling terms are swamped by the
strong interaction between polymers. Hence no major
differences in the coupled and uncoupled phase behaviour are detectable for $\chi=0.3$ (bottom panel).

\subsection{Effect of Dielectric Monomer Size \label{EffectOfF}}

\begin{figure}[tbph]
\centering
\includegraphics[width=0.5\linewidth]{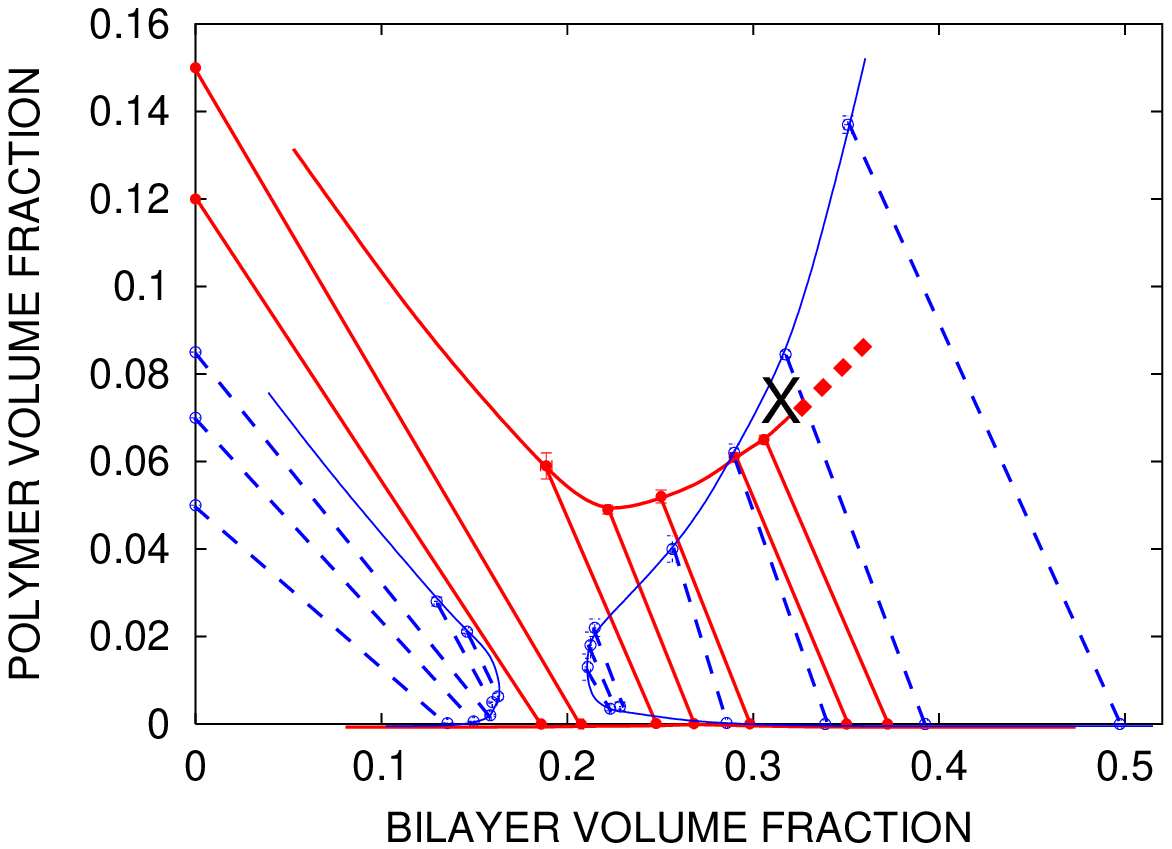}\\
\includegraphics[width=0.5\linewidth]{PaperFigures/PhaseDiagrams/s01c002phi-phi_phasediagtielined_PAPER}\\
\includegraphics[width=0.5\linewidth]{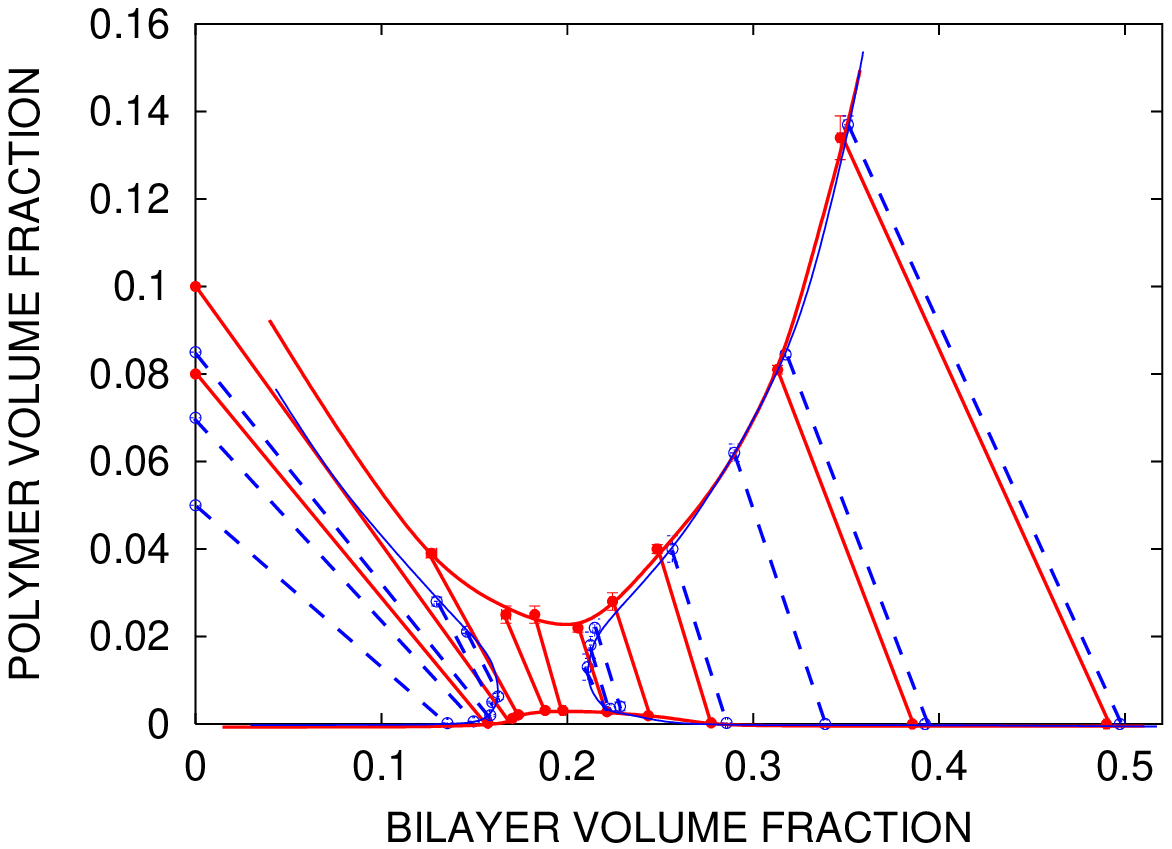}
\caption{Effect of increasing the dielectric size factor $\gamma$ on the phase behaviour of a
  polymer--doped lamellar phase in contact with a salt reservoir. Top to bottom:
  $\gamma=1.1$, $\gamma=1.5$, and $\gamma=2$. 
Method and notation as in Fig. \protect{\ref{sigma_study_fig}}.
\if{The coordinates of coexisting phases, were
  found using the graphical method discussed in the text. Coupled ({\color{red}$\bullet$})   
  and uncoupled ({\color{blue}$\circ$}) points, with associated error bars, mark these
  coordinates and have been connected by tielines (full and broken
  respectively). Approximate binodals have also been fitted through the points to
  highlight the shape of the miscibility gaps.}\fi
 For $\gamma=1.1$ in the coupled case, our method did not allow
  obtain the phase behaviour in the region of concentrated bilayers. The dotted line
  indicates what we believe to be a reasonable extrapolation of the binodal. Point $X$ is
referred to in the text.}\label{dielfact_study_fig}
\end{figure}

We now consider the role of $\gamma$, which governs the 
volume of polarisable material associated with a
lattice monomer. Shown in \fig{dielfact_study_fig} are the phase diagrams calculated when
$\gamma=1.1$,  $\gamma=1.5$ and $\gamma=2$. As
expected, an increase in $\gamma$, which implies a decrease of the amount of dielectric
material in a polymer (see \fig{lattice_figure}), reduces the difference between coupled and  uncoupled approaches. When $\gamma=1.1$ (top panel), in the bilayer dilute region
the $\rm L_\alpha L_\alpha$ coexistence predicted by the uncoupled model is replaced by a
significantly larger area of $\rm L_\alpha L$ in the coupled case. However, for
concentrated bilayer volume fractions, the coupled miscibility gap is smaller than the
uncoupled one. 

Note also that, even for $\gamma=2$ when there is a relatively small
amount of dielectric material in each lattice monomer, there is still a strong effect on
phase diagram topology near the pinch-off from one region of immiscibility to two. The phase
diagram for these parameter values is topologically sensitive to all perturbations, including
those arising from a relatively modest level of dielectric coupling between the polymeric
and electrostatic interactions.

The differences between coupled and uncoupled predictions is generally as expected, with
more dielectric material in a
monomer (small $\gamma$) giving larger enhancement 
of the miscibility gaps when coupling is added. However,
at large $\phi_b$ there appears to be an inversion in this trend (top panel), with  
the $\rm L_\alpha L_\alpha$ coexistence 
reduced with respect to the uncoupled predictions. This inversion
(mentioned already in \sect{EffSurfChargDensSect}) results from the dielectric 
screening of highly confined polymers at moderate concentrations. To see this, consider a
lamellar 
phase prepared at the composition
marked by an $X$ on the top panel of \fig{salt_study_fig}. In an uncoupled picture the
compromise between electrostatic  
repulsion and polymer attraction can be achieved through phase separation, even though,
at high concentration, the polymer attraction is of short range (the range
involves the correlation length in the solution, which falls with increasing concentration
\cite{DeGennesBook}).  
In most parts of the phase diagram, introducing coupling enhances the polymer-mediated
attraction by keeping polymers away from the high field regions close to the bilayers
(\sect{EffSurfChargDensSect}), but in this regime that tendency
is largely overcome by the high polymer osmotic pressure. Instead, the polymers between
layers help to screen the electrostatic repulsion (by reducing the dielectric constant) and
this stabilizes the lamellar phase against phase separation to a state of small $D$.

\section{Discussion \label{RelRealSystSect}}

We now discuss the relation between the phase diagrams found above and
those obtained experimentally for polymer-doped smectics. Because of the various approximations made
(including the assumptions of strongly non-adsorbing polymer and of a mean-field approach, and
neglect of all interaction terms between bilayers other than the electrostatic and polymer-mediated
forces) we focus on qualitative rather than quantitative comparisons.

\subsection{Differences Due to the Salt Reservoir}

We first must decide whether the use of a salt
reservoir (Section \ref{SemiGrand}) in our calculations affects the qualitative aspects of our predictions. 
Recall typical experiments on polymer--doped lamellar phases are performed by adding a
fixed amount of salt to each sample. When lamellar phases contact a reservoir through a
semipermeable membrane, they can expel salt into the reservoir (the Donnan effect
\cite{ZembDubois, Donnan}) reducing the repulsion between them. For a system of fixed salt
content, however, 
this expulsion is not possible. This implies a greater electrostatic repulsion between
bilayers, so that more polymer will be required to cause phase separation in a fixed salt
system. We thus expect the shape of the phase diagrams (and the tilt of the tielines) to
differ from our Semi-Grand ensemble predictions, but expect the general qualitative
features and their dependence on parameter variation to be the same.

\subsection{Comparison with Experimental Results \label{CompExpResSEct}}

Our results reproduce
several of the main qualitative features observed in experiments on polymer-doped lamellar phases,
notably $\rm L_\alpha L_\alpha$ and $\rm L_\alpha L$ phase separations (see e.g \cite{Ficheux95, Javierre01,
LigourePorte93, LigoureTheoryExp97,LigoureElec97,Ligoure2000}). However we do not
capture all the features of very dilute and very
concentrated lamellar phases. For example, an experimentally observed $\rm L_\alpha L$
coexistence involving very concentrated lamellar phases is not predicted in our phase
diagrams. Another feature which does not appear in our results is the ``closed--loop''
miscibility gap sometimes observed \cite{Ficheux95,
Javierre01}. This has been ascribed to the two different ranges of the
repulsive interactions between bilayers: the very short range of hydration and the larger
range of electrostatics \cite{Sear1}. Such repulsions compete with polymer induced
attraction at different separations between bilayers. For large bilayer volume fractions
(small separations) the phase separation terminates where the hydration force
dominates the polymer attraction. The same thing happens at small bilayer volume fractions
(large separations), where electrostatic repulsion dominates the polymer attraction. These
considerations, if Sear's model \cite{Sear1} is correct, explain why a model with no short
ranged repulsion such as ours cannot predict closed loops, but only open regions of phase
coexistence terminating in a critical point.

Our phase diagrams predict qualitatively correct trends when electrostatic parameters,
such as salt concentration and surface charge density are changed. Ligoure et al. have
investigated these changes experimentally by looking at the  phase behaviour of doped
lamellar phases of fixed polymer content. They found that progressive addition of salt
\cite{LigourePorte93, LigoureTheoryExp97} or reduction of surface charge density
\cite{LigoureElec97, Ligoure2000} cause a transition  from a single phase lamellar sample
to a $\rm L_\alpha L_\alpha$ coexistence. These qualitative trends agree with our predictions,
e.g. of a system with $\phi_b=0.2, \phi_p=0.02$ as $\sigma$ is decreased from $0.2$
(\fig{sigma_study_fig}, bottom panel) to $0.1 \ut{e\,nm^{-2}}$ (middle panel). The reduction of
the surface charge density induces the same  $\rm
L_\alpha L_\alpha$ transition as observed in experiment. 

The experimentally confirmed trends reported seem largely insensitive to the question of
whether the dielectric coupling of the polymer-mediated and electrostatic interactions is included. 
As shown above, this is mainly a quantitative rather than a qualitative difference, so long as
the evolution of phase diagrams is considered in the round. However, this does not preclude
large differences when coupled and uncoupled results are compared for particular parameter values, particularly
around the pinch-off point where the immiscible region splits into two pieces (Fig.\ref{sigma_study_fig}, etc.).

\section{Conclusions \label{ModIIConclCommSummSect}}

Our results show that the dielectric proprieties of uncharged polymers confined between
charged surfaces couple together electrostatic and polymeric effects in a
non-additive fashion, which can have significant quantitative impact on phase diagrams. 
Since the monomers in a chain are electrically
polarised by the (nonuniform) electric field of the double layer, the depletion of polymers
close to the surface is electrostatically enhanced. Conversely, the polarisation of
the monomers provides an additional source of screening of the
electrostatics. Consequently, the osmotic
interaction between 
surfaces as a function of distance is also affected: the range of the electrostatic 
part of the interaction is effectively reduced with respect to a model which
ignores coupling, whereas that of the depletion attraction is effectively increased.

Our results, based on a mean-field theory that couples both sorts of interaction, gives
a good account of the phase behaviour of
polymer-doped lamellar phases. Our study of its evolution
under parameter variation confirms that the coupling effects do not drastically alter
the trends from the uncoupled case but do alter the quantitative predictions; near topological changes 
in phase diagram, but not elsewhere, these quantitative corrections change the qualitative behaviour.
Coupled and uncoupled predictions differ most under the following conditions (i) moderately high surface charge
densities (since strong 
polarisation  of the polymers requires high fields); (ii) marginal solvency, i.e., near-{\it theta} conditions (where the response in polymer concentration to external energy shifts is maximized); and (iii) polymers containing a
large volume of low dielectric material (for obvious reasons of coupling strength). 
On the other hand, coupling effects were not discernible for
lamellar phases with small surface charges, those doped with polymers residing in a good
solvent, or chains with insufficient dielectric contrast.

Future experimental investigations could perhaps put these coupling effects more strongly in evidence.
For example, it would be interesting to compare the phase behaviour of lamellar phases doped with polymers
of strongly contrasting dielectric properties (changing either the monomeric dielectric contrast and/or
dielectric volume) under matched conditions of near-{\it theta} solvency. 
Alternatively, instead of investigating the phase behaviour, a more direct test of the interplay between
polymeric and electrostatic interactions could
be performed by directly measuring the force between charged surfaces across a polymer solution using a surface force apparatus, as in \cite{KekicheffPEG}.
From a theoretical point of view, it might also be of interest to consider the
consequences of dielectric coupling for the interaction between curved surfaces, such as
those of colloids (within the Derjaguin approximation \cite{Safran} it is easy
to apply our model to this situation), as has been done by Borukhov et
al. \cite{BorukhovAndelman} in the case of charged surfaces interacting across a solution
of polyelectrolytes. It might also be of interest to consider kinetic processes,
where coupling effects should be less subtle, such as the diffusion of a dielectric
polymer onto a charged surface. 
More generally we hope our work will stimulate further experiments
on the nonadditivity of electrostatic and depletion interactions in the
presence of dielectric contrast.  

\subsection*{Acknowledgement}
This work was funded in part under EPSRC Grant GR/S10377/01.

\begin{appendix}
\section{Derivation of Model Equations and Osmotic Pressure \label{Appdx}}

Equations \eq{PBPolyMono} and \eq{PolyPolyEq0} follow from the null variations of the free
energy action \eqn{FEAP}. For this purpose it is convenient to write \eqn{FEAP} fully:
\begin{eqnarray}\label{FEAP1}
\lefteqn{\mathcal{F}_{\Omega}[V, n_i, \phi]=\int_\Gamma f_{\Omega}(V, n_i,
\phi)\,d\vect{r} = }\nonumber\\  & =
&\int_\Gamma\left(-\frac{1}{2}\epseff(\phi)(\nabla V)^2 + \sum_{i=+,-} n_i q_i V+
T \sum_{i=+,-} n_i (\ln{n_i}-1)  -\sum_{i=+,-}
\mu_i n_i \,+ \right.\\  & & \left. {}+ \frac{T}{a^3}\left
[\frac{\phi}{N}(\ln{\frac{\phi}{N}}-1)+(1-\phi)\ln(1-\phi)+\chi \phi (1-\phi)
  -\frac{\mu_p}{T}\right] +\,
\frac{T}{36 a} \frac{(\nabla \phi)^2}{\phi} \right)\,d\vect{r} \nonumber
\end{eqnarray}
Note that the ideal contribution to the polymer free energy differs here from the
standard Flory--Huggins expression by an extra term $\phi/N$ (we follow the convention of
\cite{CatesPagonabarraga}). The term is linear in $\phi$, and so is inconsequential to any
derived quantity.

\subsection{The Modified Poisson--Boltzmann Equation}

Performing a variation with respect to the electrostatic potential, we set:
$\minifderiv{\Action}{V} = 0$ or, equivalently, $\frac { \partial \action }{ \partial V
}-\nabla{ \frac { \partial \action }{\partial \nabla V} } = 0$. From this we recover the
Maxwell equation:
\begin{equation}\label{PolyPoisson0}
\nabla(\epseff(\phi)\nabla V)=-\sum_{i=+,-}n_iq_i
\end{equation}
where $\epseff(\phi)$ is given by \eqn{MGphipoly}.

Similarly a variation with respect to the ion number density $\minifderiv{\Action}{n_j} =
0$, i.e. $\pderiv{\action}{n_j}-\nabla\pderiv{\action}{\nabla n_j} = 0$ yields the
Boltzmann factor for ion concentrations:
\begin{equation}\label{PolyBoltzIons0}
n_i=n_i^r \exp{-q_iV/T}
\end{equation}
where, $n_i^r\equiv\exp{-\mu_i/T}$ defines the chemical potential $\mu_i$ of ion $i$,
fixed by the reservoir. All symbols in \eq{PolyPoisson0} and \eq{PolyBoltzIons0} have also
been previously defined.
Substituting \eqn{PolyBoltzIons0} into \eq{PolyPoisson0} yields a modified
Poisson--Boltzmann equation in $V$:  
\begin{equation}\label{PolyPB0}
\nabla(\epseff(\phi)\nabla V)=-\sum_{i=+,-}n_i^r q_i\exp{-q_iV/T}
\end{equation}
which, in the case of monovalent ions, becomes \eq{PBPolyMono}.

\subsection{The Polymer Equation}

A variation with respect to the polymer volume fraction
$\minifderiv{\Action}{\phi} = 0$,
i.e. $\pderiv{\action}{\phi}-\nabla\pderiv{\action}{\nabla \phi} = 0$, yields:
\begin{eqnarray}\label{Polymer0}
-\frac{1}{2}\deriv{\epseff}{\phi}(\nabla V)^2 &+&
\frac{T}{a^3}\left[\frac{1}{N}\ln{\frac{\phi}{N}}-\ln(1-\phi)-1+ \chi -2\chi
  \phi-\frac{\mu_p}{T}\right]+\nonumber\\ & & -
\frac{T}{36a}\left[2\frac{\nabla^2\phi}{\phi}-\left(\frac{\nabla\phi}{\phi}\right)^2\right]=0
\end{eqnarray}
This describes the response of the polymer concentration to external
perturbations such as electric fields and confining hard boundaries. When such
perturbations are not present, we recover a uniform polymer solution (the
reservoir). In this case $\phi=\phi^r$, and \eqn{Polymer0} fixes
the chemical potential of the solution:
\begin{equation}\label{PolChemPot}
\frac{\mu_p}{T}= \left[\frac{1}{N}\ln{\frac{\phi^r}{N}}-\ln(1-\phi^r)-1+\chi -2\chi
\phi^r\right]
\end{equation}
To simplify \eqn{Polymer0} we change
variable to $\psi\equiv\phi^{1/2}$. This 
implies \ensuremath{2\nabla^2\phi/\phi-(\nabla\phi/\phi)^2=4\nabla^2\psi/\psi}. Thus,
substituting \eq{PolChemPot} into \eq{Polymer0} and carrying out the differentiation of
the Maxwell--Garnett relation \eq{MGphipoly} for \epseff we obtain:
\begin{eqnarray*}\label{PolymerPsi0}
\frac{1}{2}\frac{3K \alpha\epsilon_1}{(1+K \alpha\psi^2)^2}(\nabla V)^2 &+& \frac{T}{a^3}\left[
\frac{1}{N}\ln{\left(\frac{\psi^2}{\psi^{r\,2}}\right)}- \ln\left(
\frac{1-\psi^2}{1-\psi^{r\,2}} \right)- 2\chi(\psi^2-\psi^{r\,2}) \right]+ \nonumber\\ & &
- \frac{T}{9a}\frac{\nabla^2\psi}{\psi}=0
\end{eqnarray*}
which is easily rearranged into \eqn{PolyPolyEq0}.

\subsection{Osmotic Pressure from a Conservation Law}

In mechanics, when the Lagrangian (density) does not explicitly depend on time, the
Hamiltonian is a constant of the motion \cite{Schwinger}. Similarly, the
free energy action density of our model does not explicitly depend on position; we can
thus define the following conserved quantity: 
\begin{equation}\label{OsmPiModIIEqn}
{\cal H}\equiv\sum_m p_m \nabla Q_m - \action= const.
\end{equation}
where $\action$ is the free energy action density defined by \eq{FEAP1}, and $Q_m$ and
$p_m$ are, respectively, the generalised coordinates and
momenta of the problem. The generalised momenta, defined as $p_m \equiv \pderiv{\action} {\nabla
  Q_m}$, are:
\begin{eqnarray}
p_{n_i}&=&\pderiv{\action}{\nabla n_i}=0\\ p_{\phi}&=&\pderiv{\action}{\nabla
\phi}=\frac{2 T}{36 a}\frac{\nabla{\phi}}{\phi}\\ p_{V}&=&\pderiv{\action}{\nabla
V}=-\epsilon(\phi_d) \nabla V
\end{eqnarray}
so that by \eq{OsmPiModIIEqn} we can define the following quantity:
\begin{eqnarray}\label{PolyPi1}
\lefteqn{ {\cal H} = p_{V} \nabla V + p_{\phi} \nabla \phi - \action = }\nonumber\\  & = &-
\frac{1}{2}\epseff(\phi)(\nabla V)^2 - \sum_{i=+,-} n_i q_i V - T \sum_{i=+,-} n_i
(\ln{n_i}-1) + \sum_{i=+,-} \mu_i n_i + \nonumber\\  & - & \frac{T}{a^3}\left[
\frac{\phi}{N}(\ln{\frac{\phi}{N}}-1)+(1-\phi)\ln(1-\phi)+\chi \phi (1-\phi)
-\frac{\mu_p}{T} \right] + \frac{T}{36 a} \frac{(\nabla \phi)^2}{\phi}\nonumber= const.
\end{eqnarray}
Upon substitution of \eq{PolChemPot} for the polymer chemical potential and
\eq{PolyBoltzIons0} for the ion densities, \eq{PolyPi1} becomes, after a few algebraic
manipulations:
\begin{eqnarray}\label{PolyPiGen0}
{\cal H} & = & - \frac{1}{2}\epseff(\phi)(\nabla V)^2  + T \sum_{i=+,-} n^r_i \exp{-q_iV/T} +
\\  & - & \frac{T}{a^3}\left[
\frac{\phi}{N}(\ln{\frac{\phi}{\phi^r}}-1)-\phi\ln\left(\frac{1-\phi}{1-\phi^r}\right)
+\chi \phi (2\phi^r-\phi) +\phi+\ln(1-\phi) \right] + \nonumber\\  & + & \frac{T}{36 a}
\frac{(\nabla \phi)^2}{\phi}= const.\nonumber
\end{eqnarray}
The total stress on a surface in an ionic solution in the vicinity of a charged surface
has osmotic and electrostatic contributions which can be calculated from a a stress tensor
(the Maxwell stress tensor with an added isotropic osmotic contribution, see Podgornik in 
\cite{ElecEffSoftMatt}). The normal component of this stress, $\Sigma_{xx}$ is identical with the first
two terms, the electrostatic and ionic contributions respectively, of
\eqn{PolyPiGen0}. The other terms represent the contributions to $\Sigma_{xx}$ due to the
polymer solution. Hence we identify ${\cal H}$ with $\Sigma_{xx}$. On the midplane ($x = D/2$) the
electrostatic contribution vanishes by symmetry, as do all gradient terms. Here (only) the conserved
quantity $\Sigma_{xx}$ can be identified with the osmotic pressure $\Pi$, which has ionic and
polymeric contributions.

\subsubsection*{Reservoir Pressure and Net Osmotic Pressure}

In the reservoir we have an ideal ionic solution mixed with a Flory--Huggins polymer solution, and no
net field contributions. When evaluated in the ``reservoir limit'' (as we did when
deriving the chemical potential), we thus expect the mid-plane osmotic pressure \eq{PolyPiGen0} to
reduce to the sum of Van't Hoff ideal contribution for the ions and a Flory--Huggins pressure term
for the polymer solution. Setting $V=0=\nabla V$ and $\phi=\phi^r$, \eq{PolyPiGen0}
becomes:
\begin{equation}\label{PolyPiFH}
\Pi^r  =   T \sum_{i=+,-} n^r_i + \frac{T}{a^3}\left[ \frac{\phi^r}{N} -\chi \phi^{r\,2}
  -\phi^r -\ln(1-\phi^r) \right]
\end{equation}
\eqn{PolyPiFH} agrees with our expectations and is an expression for the reservoir
pressure. It also provides a check of our derivation of $\Sigma_{xx}$ as a
conserved quantity, since the polymer contribution to \eqn{PolyPiFH} can be independently
and directly derived by differentiation of the Flory--Huggins free energy of a bulk polymer solution
(e.g.: see section III.1.3 of Reference \cite{DeGennesBook}).

To find the net force per unit area between the plates, we need to subtract the reservoir
osmotic pressure \eq{PolyPiFH} from the midplane osmotic pressure $\Pi$ which is numerically
equal to the conserved quantity defined in \eq{PolyPiGen0}:
\begin{eqnarray}\label{PolyPiNet0}
\lefteqn{\Pi^{net}=\Pi - \Pi^r }\nonumber\\ & &  = -\frac{1}{2}\epseff(\phi)(\nabla V)^2
+ T \sum_{i=+,-} n^r_i (\exp{-q_iV/T}-1) +   \\  & - &  \frac{T}{a^3}\left[
\frac{\phi}{N}\ln{\frac{\phi}{\phi^r}}+\frac{(\phi^r-\phi)}{N}+
(1-\phi)\ln\left(\frac{1-\phi}{1-\phi^r}\right)- \chi (\phi^r-\phi)^2-
(\phi^r-\phi)\right] + \nonumber\\ & + & \frac{T}{36 a} \frac{(\nabla \phi)^2}{\phi}=
const.\nonumber
\end{eqnarray}
In principle the net force per unit area could be found by evaluating the (conserved) right hand side at
at any position $x$ between the plates. If we assume monovalent ions,
\eq{PolyPiNet0} becomes:
\begin{eqnarray}\label{PolyPiMono}
\lefteqn{ \Pi^{net} =  -\frac{1}{2}\epseff(\phi)(\nabla V)^2  + 4 n^r_s T
\sinh^2\left(\frac{eV}{2T}\right) + }\\ & & -
\frac{T}{a^3}\left[\frac{\phi}{N}\ln{\frac{\phi}{\phi^r}}+\frac{(\phi^r-\phi)}{N}+
(1-\phi)\ln\left(\frac{1-\phi}{1-\phi^r}\right) - \chi (\phi^r-\phi)^2-
(\phi^r-\phi)\right] + \nonumber\\ & & +   \frac{T}{36 a} \frac{(\nabla
\phi)^2}{\phi}=const.\nonumber
\end{eqnarray}
For the case under study of opposing flat
surfaces, the electric field and all gradient terms vanish
at the midplane by symmetry; choosing this as the place to evaluate the right hand side, \eq{PolyPiMono}
reduces to \eqn{PolyPiMono0}.
\end{appendix}


\begin{thebibliography}{10}
\bibitem{NOTE_prefactor_SG}
Note that some authors use a coefficient where the numerical factor $36$ in the
  denominator of the SG coefficient is replaced by $24$. This corresponds to a
  use of the small wavelength limit of the free energy expansion in the RPA
  \cite{CatesPagonabarraga}.
\bibitem{NOTE_approx_sat}
This choice of parameters entails that the condition $\lambda\gg R \gg a
  \phi^{-1/3}$ discussed in \sect{VarFormSect} is marginally satisfied at best.
\end{thebibliography}

\providecommand{\refin}[1]{\\ \textbf{Referenced in:} #1}

\end{document}